\documentclass[lettersize,journal]{IEEEtran}
\usepackage{amsmath,amsfonts}
\usepackage{amsthm}
\usepackage{algorithmic}
\usepackage[ruled,vlined]{algorithm2e}
\usepackage{array}
\usepackage[caption=false,font=normalsize,labelfont=sf,textfont=sf]{subfig}
\usepackage{textcomp}
\usepackage{multirow}
\usepackage{threeparttable}
\usepackage{bm}
\usepackage{setspace}
\usepackage{bbding}
\usepackage{pifont}
\usepackage{stfloats}
\usepackage{subfloat}
\usepackage{url}
\usepackage{verbatim}
\usepackage{makecell}
\usepackage{graphicx}
\usepackage{booktabs}
\usepackage{cite}
\usepackage{tikz}
\usepackage{hyperref}
\hypersetup{
    colorlinks=true,
    urlcolor=blue,    
    linkcolor=black,  
    citecolor=black   
}
\usepackage[hang,flushmargin]{footmisc}

\newcommand{\down}{\vspace*{0.05in}}

\usepackage{xcolor} 
\usepackage{amsthm} 
\newtheoremstyle{example}
  {3pt} 
  {3pt} 
  {} 
  {} 
  {\bfseries} 
  {.} 
  {.5em} 
  {} 

\theoremstyle{example}

\usepackage{enumitem}
\setlist[itemize]{leftmargin=*}

\usepackage{xcolor} 
\usepackage{amsthm} 
\newtheoremstyle{Definition}
  {3pt} 
  {3pt} 
  {} 
  {} 
  {\bfseries} 
  {.} 
  {.5em} 
  {} 

\theoremstyle{Definition}

\usepackage{xcolor} 
\usepackage{amsthm} 
\newtheoremstyle{Property}
  {3pt} 
  {3pt} 
  {} 
  {} 
  {\bfseries} 
  {.} 
  {.5em} 
  {} 

\theoremstyle{Property}

\usepackage{xcolor} 
\usepackage{amsthm} 
\newtheoremstyle{Proposition}
  {3pt} 
  {3pt} 
  {} 
  {} 
  {\bfseries} 
  {.} 
  {.5em} 
  {} 

\theoremstyle{Proposition}

\usepackage{xcolor} 
\usepackage{amsthm} 
\newtheoremstyle{Problem}
  {3pt} 
  {3pt} 
  {} 
  {} 
  {\bfseries} 
  {.} 
  {.5em} 
  {} 

\theoremstyle{Problem}

\begin{document}

\title{Stellar: Scalable Multimodal Document Retrieval for Natural Language Queries}

\author{Yuxiang Guo,
        Zhonghao Hu,
        Yuren Mao,
        Yuhang Liu,
        Congcong Ge,
        Xiaolu Zhang,\\
        Jun Zhou,
        and Yunjun~Gao, ~\IEEEmembership{Senior Member,~IEEE}
       
\thanks{Y. Guo, Z. Hu, Y. Mao, Y. Liu, C. Ge, and Y. Gao are with Zhejiang University, Hangzhou 310027, China (e-mail: guoyx@zju.edu.cn; zhonghao.hu@zju.edu.cn; yuren.mao@zju.edu.cn; lyh65535@zju.edu.cn; gcc@zju.edu.cn; gaoyj@zju.edu.cn).}
\thanks{X. Zhang and J. Zhou are with Ant Group, Hangzhou 310013, China (e-mail:  yueyin.zxl@antfin.com; jun.zhoujun@antfin.com)}
}


\markboth{IEEE Transactions on Knowledge and Data Engineering,~Vol.~XX, No.~XX, XXX~XXXX}{Li \MakeLowercase{\textit{et al.}}:Snoopy: Effective and Efficient Semantic Join
Discovery via Proxy Columns}


\maketitle

\begin{abstract}
Multimodal document retrieval---selecting the most relevant multimodal document from a large corpus to answer a natural language
query---plays an essential role in Retrieval-Augmented Generation (RAG) systems. 
State-of-the-art methods represent each document and query with multiple token-level embeddings and use late interaction to achieve high effectiveness. However, such multi-vector representations incur substantial memory overhead during retrieval, leading to poor scalability and hindering real-world deployment. In this paper, we present \textsc{Stellar}, a scalable multimodal document retrieval framework that stores token-level document embeddings on disk and loads only a small set of candidate embeddings into memory for late interaction. \textsc{Stellar} comprises two key components: (i) Lexical Representation-based Filtering (LRF), which fine-tunes a Multimodal Large Language Model (MLLM) as a sparse encoder to produce high-quality lexical representations, enabling efficient and effective document filtering to significantly reduce the candidate set; (ii) Efficient Disk-backed Late Interaction (DLI), which designs an on-disk token embedding storage layout guided by a balanced clustering algorithm, and dynamically loads only the necessary token embeddings into memory using a simple yet effective cost model. Extensive experiments on four real-world benchmarks and a newly presented large-scale dataset demonstrate that \textsc{Stellar} reduces memory overhead and query latency by 1-2 orders of magnitude compared to existing methods without compromising retrieval effectiveness. 
\end{abstract}


\section{Introduction}
\label{sec:intro}
Large language models (LLMs)~\cite{gpt32020,gpt4ocard} have demonstrated strong performance across diverse tasks, yet hallucination remains a major challenge for their reliable deployment in domain-specific applications~\cite{HallucinationL25}. Retrieval-Augmented Generation (RAG)~\cite{rag} mitigates this issue by enabling LLMs to incorporate external knowledge through information retrieval. While most existing RAG systems operate over text corpora~\cite{Self-RAG,FIT-RAG,QAEA-DR}, in practice, knowledge is often embedded in multimodal documents that integrate text, charts, and other visual elements~\cite{DocDB,VisRAG2025}.  This motivates the need for multimodal document retrieval to locate relevant documents to answer users' natural language (NL) queries. 

\begin{figure}
	\centering
    \includegraphics[width=1\linewidth]{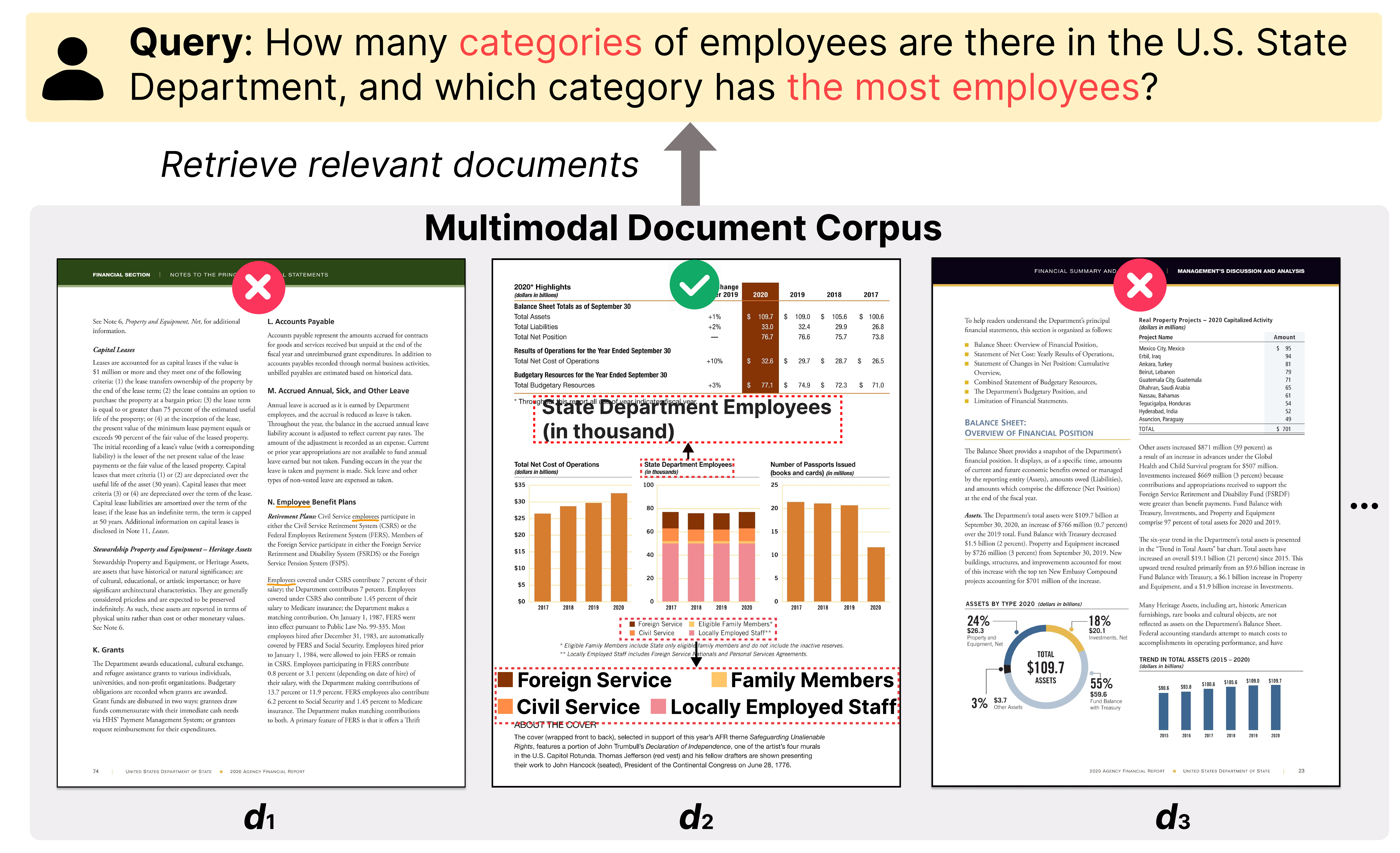}
    \vspace{-4mm}
    \caption{Multimodal document retrieval for NL queries.}
    \label{fig:exp1}
    \vspace{-3mm}
\end{figure}

As an example shown in Figure~\ref{fig:exp1}, given a user query, multimodal document retrieval aims to identify the document $d_2$ from the large corpus,  as it contains the information about ``employee categories'' and ``employee counts'' represented by legends and bar charts, which can be used to answer the query. 
Once $d_2$ is retrieved, it can be provided to a generator to generate a response. The quality of downstream question answering depends on the retrieval step.

\begin{figure}
	\centering
    \includegraphics[width=1\linewidth]{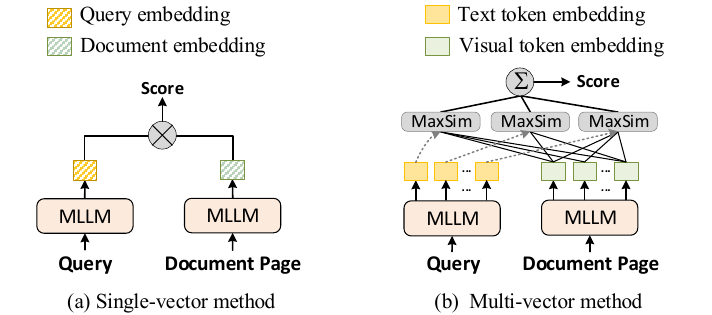}
    \vspace{-4mm}
    \caption{An illustration of the single-vector-based method and multi-vector method with late interaction.}
    \vspace{-3mm}
    \label{fig:exp2}
\end{figure}

Traditionally, multimodal document retrieval relies on Optical Character Recognition (OCR)~\cite{Shi_2016_CVPR} to extract text, followed by standard text-based retrieval methods~\cite{bm2509,bgelarge2023}. Such pipelines often suffer from limited effectiveness due to OCR errors and the loss of visual information.
With the rapid development of Multimodal Large Language Models (MLLMs)~\cite{mllm}, recent studies~\cite{colpali2025, VisRAG2025} have shifted toward vision-centric paradigms: each document page is rendered as an image and directly encoded by MLLMs. 
As shown in Figure~\ref{fig:exp2}(a), single-vector-based methods~\cite{VisRAG2025, DSE24} encode each document page into a single dense vector; however, such a global representation often fails to capture the fine-grained semantics of complex documents. To address this, multi-vector methods~\cite{colpali2025,M3DocRAG2024} encode both queries and documents into multiple token-level embeddings, as shown in Figure~\ref{fig:exp2}(b), and further employ the late interaction mechanism~\cite{ColBERT2020}, where each query token interacts with all document
tokens to compute the overall query-document relevance. 


Although multi-vector methods have achieved state-of-the-art (SOTA) retrieval effectiveness~\cite{colpali2025,M3DocRAG2024}, they suffer from poor scalability. 
At retrieval time, they require all token-level document embeddings to reside in memory and perform late interaction, incurring substantial computation and memory overhead.
For example, the SOTA method ColPali~\cite{colpali2025} is estimated to require over 800 GB of memory to store token embeddings for a 2.4-million-document corpus~\cite{docmatix}.
This results in considerable deployment costs, exceeding \$8,500 per month
on a standard cloud instance (e.g., AWS r7iz.32xlarge\footnote{https://instances.vantage.sh/aws/ec2/r7iz.32xlarge?currency=USD}). Although quantization-based approximate search methods~\cite{PLAID,ColBERTv2} can reduce the memory footprint of token-level embeddings by compressing them offline and keeping the quantized representations resident in memory during retrieval, this design does not fundamentally address the scalability challenge. 
First, the offline compression stage requires loading all original token embeddings into memory, resulting in the same peak memory footprint as the uncompressed representations. Second, retrieval-time memory usage still grows rapidly with corpus size, incurring substantial memory footprints at scale.
This motivates a paradigm shift: instead of compressing token embeddings to fit in memory, \textit{we store original token embeddings on inexpensive disk and selectively load only a small subset into memory for late interaction}. 
However, introducing disk-backed late interaction raises two new challenges.

\down\noindent {\textbf{Challenge I:}} \textit{How to design an effective and lightweight filter?}
In a disk-based retrieval paradigm,
loading all documents' token embeddings from disk into memory for late interaction is impractical, as it would undermine the purpose of memory saving. Therefore, 
an effective filtering stage is essential to identify a small candidate set. 
However, designing such a filter is non-trivial, as it must balance efficiency and recall: the filter should be lightweight to avoid introducing substantial memory overhead or latency to the overall framework, yet powerful enough to ensure that relevant candidates are not prematurely discarded.
Existing solutions typically employ dense representations to retrieve top-ranked candidates~\cite{DSE24,PLAID}. While effective, the high-dimensional dense embeddings incur substantial storage overhead and become a major bottleneck for end-to-end retrieval at large corpus scales (see Section~\ref{subsec:rq3}).

\down\noindent {\textbf{Challenge II:}} \textit{How to reduce disk-to-memory data loading time?} 
After candidate documents are identified, their token-level embeddings must be loaded into memory for late interaction, making disk-to-memory data loading a critical efficiency bottleneck due to I/O latency.
This challenge is twofold, involving both offline data organization and online loading strategy.
First, without careful layout design, token embeddings of semantically related documents are stored non-contiguously across disk blocks, resulting in expensive random I/O.
Second, even with an optimized storage layout, the system faces a dynamic online loading dilemma: sequentially loading entire block  embeddings improves I/O efficiency but introduces overhead from non-candidate embeddings, while selectively loading only candidate embeddings minimizes data transfer but triggers random I/O. 
Consequently, a one-size-fits-all strategy would inevitably lead to suboptimal performance.

To surmount these challenges, we present \textsc{Stellar}, a \underline{s}calable re\underline{t}ri\underline{e}va\underline{l} framework via efficient
fi\underline{l}tering and disk-backed l\underline{a}te inte\underline{r}action.
To address \textbf{Challenge I}, 
we design a siamese sparse dual-encoder based on a pre-trained MLLM, repurposing its next-token prediction head as a shared projection layer that maps both documents and queries into the model's vocabulary space, followed by a sparsification step that retains only the most informative dimensions. The resulting sparse representations support effective and efficient document filtering via an in-memory inverted index.
To tackle \textbf{Challenge II}, we propose a balanced clustering algorithm that groups semantically related documents based on the learned sparse lexical representations. The token embeddings of all documents within each cluster are stored contiguously in a single disk block, preserving semantic locality while maintaining balanced block sizes for stable I/O performance. Building on this layout, we further introduce a cost-aware loading strategy guided by a simple yet effective cost model, which adaptively decides whether to load entire blocks or only the required token embeddings, thereby optimizing data loading efficiency. 
Our main contributions are as follows: 
 
\begin{itemize}[leftmargin=*]
\item \textbf{Scalable retrieval framework.} 
We introduce \textsc{Stellar}, a scalable framework for multimodal document retrieval that 
addresses the memory-intensive limitations of existing approaches through a representation-storage co-design. 

\item \textbf{MLLM-based lexical representation.}
We propose a novel lexical representation learning method that repurposes the pre-trained prediction head of MLLMs to project complex multimodal document pages into a sparse lexical space, enabling effective and efficient document filtering.

\item \textbf{Efficient disk-to-memory loading.} We design a balanced clustering-based on-disk layout, and a cost-aware data loading strategy guided by a simple yet effective cost model, substantially reducing I/O overhead and enabling low-latency retrieval. 

\item \textbf{Large-scale benchmark.} Since no public large-scale multimodal document retrieval dataset is available, 
we construct LargeDoc, a large-scale benchmark dataset, and publicly release it for future research. 
 
\item \textbf{Extensive experiments.} Our extensive experiments on five real-world datasets show that \textsc{Stellar} reduces memory usage and query latency by 1-2 orders of magnitude compared with existing multi-vector methods, while maintaining state-of-the-art effectiveness.
\end{itemize}

\noindent \textbf{Roadmap.} The remainder of this paper is organized as
follows. Section~\ref{sec:relatedwork}  reviews related works. Section~\ref{sec:pre} provides the preliminaries. Section~\ref{sec:framework}
overviews the framework \textsc{Stellar}. Section~\ref{sec:lrf} and Section~\ref{sec:dli}
introduces two key components of \textsc{Stellar}. Section~\ref{sec:exp} reports
experimental results and our findings. Section~\ref{sec:conclusion} concludes this paper. 
\section{Related Work}
\label{sec:relatedwork}
\subsection{Multimodal Document Retrieval} 
Multimodal retrieval~\cite{WangOYZZ14,ErfanianJA24,LiYKWLX24,MUST} is a foundational task aimed at bridging the semantic gap between disparate modalities. Early paradigms predominantly focus on cross-modal alignment, where models like CLIP~\cite{clip21}, BLIP~\cite{blip22023}, and SigLIP~\cite{siglip23}  leverage contrastive learning~\cite{he2020moco} to map images and short text into a unified latent space. While effective for cross-modal matching, these methods often falter when applied to multimodal document retrieval. 
Unlike cross-modal retrieval (e.g., retrieve images by text), each multimodal document page encompasses elements of multiple modalities and complex layout structures. As highlighted in~\cite{DSE24}, CLIP-style encoders are suboptimal for complex multimodal document representation. With the development of powerful Multimodal Large Language Models (MLLMs), recent studies~\cite{VisRAG2025, DSE24, colpali2025} try to render each document page as an image and leverage MLLMs to generate high-quality embeddings for dense similarity search, and achieve promising effectiveness.

\subsection{Multi-Vector Late Interaction}
Single-vector representations are often insufficient for capturing the rich semantics of complex documents. To address this, multi-vector late-interaction methods~\cite{ColBERT2020} are proposed to encode both documents and queries as sets of token-level embeddings, and compute relevance between token pairs. Representative approaches include ColPali~\cite{colpali2025} for multimodal document retrieval and M3DocRAG~\cite{M3DocRAG2024}, which adopts ColPali as its retriever.
Despite their effectiveness, such approaches incur substantial computational and memory overhead, as they require storing and comparing large numbers of token embeddings.

While quantization-based approximate search methods~\cite{PLAID,ColBERTv2} can compress token embeddings into centroids and quantized residuals, they do not fundamentally resolve the scalability challenge and typically trade retrieval effectiveness for efficiency. In contrast, widely adopted approximate nearest neighbor (ANN) search techniques, 
including in-memory approaches such as HNSW~\cite{HNSW}, and disk-based methods such as DiskANN~\cite{diskann} and Starling~\cite{Starling}, are primarily designed for single-vector retrieval and are not directly compatible with multi-vector late interaction. Although some existing vector search systems (e.g., Milvus~\cite{Milvus}), extend single-vector retrieval to multi-vector settings, they merely aggregate similarity scores from isolated single-vector searches---a design that fundamentally diverges from the token-level late interaction paradigm prevalent in multimodal document retrieval.

\subsection{Sparse Retrieval}
Sparse retrieval traditionally relies on literal term matching within a shared vocabulary space to enable efficient candidate filtering via inverted indexes. Early approaches such as BM25~\cite{bm2509} calculate relevance scores based on term-frequency and inverse document-frequency statistics. While highly efficient, these methods suffer from the lexical mismatch as they cannot capture semantic synonyms.  To address this, recent neural sparse retrievers like SPLADE~\cite{SPLADE} leverage BERT~\cite{bert} to jointly learn term expansion and term weighting, substantially improving retrieval effectiveness while preserving sparsity.  This paradigm has recently been extended to the multimodal domain. For instance, STAIR~\cite{stair} and MLSR~\cite{mlsr} map images and text into sparse token spaces for cross-modal tasks. However, these methods typically use small encoder-only models and are ill-suited for complex multimodal document pages. In contrast, \textsc{Stellar} bridges this gap by introducing the powerful MLLM, and adapts the decoder as a sparse encoder for document filtering.  
\section{Preliminaries}
\label{sec:pre}
In this section, we first provide the problem statement, and then introduce the backgrounds of multimodal large language models and late interaction mechanism.

\subsection{ Problem Statement}
Let $d$ denote a multimodal document page, as exemplified in Figure~\ref{fig:exp1}, which comprises both textual content and visual elements. Following previous studies~\cite{colpali2025,VisRAG2025}, 
we represent each document page as a rendered image, referred to as a document image or document screenshot. This rendered page image serves as the atomic unit of retrieval. 
In this paper, we use the terms ``document'' and ``page'' interchangeably.
A multimodal document corpus $D = \{d_1,  \dots, d_{|D|}\}$ is a set of such documents.

Given a multimodal document corpus $D$ and a natural language (NL) query $q$, multimodal document retrieval aims to identify query-relevant documents that contain the answer to the query $q$. This relevance is determined using a similarity metric $\operatorname{sim}(q, d)$.
Following previous studies~\cite{birdie,DSE24}, we focus on single-document retrieval. We leave multi-document retrieval, in which answers are distributed across various multimodal documents, as future work.

\subsection{Multimodal Large Language Models}
\label{sec:pre_mllm}
An MLLM typically consists of a vision encoder and an LLM decoder with a pre-defined vocabulary $\mathcal{V}$. 
Given a visual input (e..g, a document image), the MLLM first divides it into a sequence of fixed-size patches. 
These patches are processed by the vision encoder to extract visual features, which are then mapped into the language embedding space via a learnable adapter to produce visual tokens: $\{\bm{t}_1^{\text{vis}},  \dots, \bm{t}_m^{\text{vis}}\}$,
where $m$ is the number of patches, which depends on the resolution of the input image and the patch size.
For a textual input, such as a query $q$, 
the LLM's tokenizer first yields a sequence of discrete token indices, which are subsequently transformed into textual tokens:
$
\{\bm{t}_1^{\text{text}},  \dots, \bm{t}_n^{\text{text}}\}$, where $n$ is the number of tokens.
The sequences from both modalities are concatenated to form a unified input sequence, which is sent to the LLM decoder to compute hidden states.
Finally, the LLM autoregressively generates textual output by projecting the hidden states through a language modeling head \( f_{\text{head}} \) to compute a probability distribution over the vocabulary $\mathcal{V}$.

\subsection{Late Interaction Mechanism}
To capture fine-grained semantic alignments between queries and documents, multi-vector representation learning~\cite{colpali2025} avoids compressing an entire document into a single vector. Instead, it encodes each document $d$ and query $q$ into a sequence of token-level embeddings, denoted as $ \{\bm{e}^d_1, \dots, \bm{e}^d_m\}$ and $ \{\bm{e}^q_1, \dots, \bm{e}^q_n\}$, respectively. Here, $m$ and $n$ represent the number of visual tokens (e.g., image patches) in document $d$ and text tokens in query $q$. Based on these representations, the late interaction mechanism~\cite{ColBERT2020} employs a \textsf{MaxSim} operator to compute the relevance score. This mechanism allows each query token to independently align with the most semantically similar part of the document, thereby preserving fine-grained semantic relations. The similarity score is formally defined as:
\begin{equation}
\label{eq:late_interaction}
\operatorname{score}_\text{mul}(q,d)=\sum_{i=1}^{n} \max _{j=1}^{m}\left(\mathbf{e}_i^{q} \cdot \mathbf{e}_j^{d}\right)
\end{equation}
where $\mathbf{e}_i^{q}$ and $\mathbf{e}_j^{d}$ denote the $L_2$-normalized embeddings of the $i$-th query token and the $j$-th document token, respectively; $n$ and $m$ are the corresponding token counts; and $(\cdot)$ denotes the inner product.

\section{Overview of \textsc{Stellar}}
\label{sec:framework}
\begin{figure*}
	\centering
    \includegraphics[width=1\linewidth]{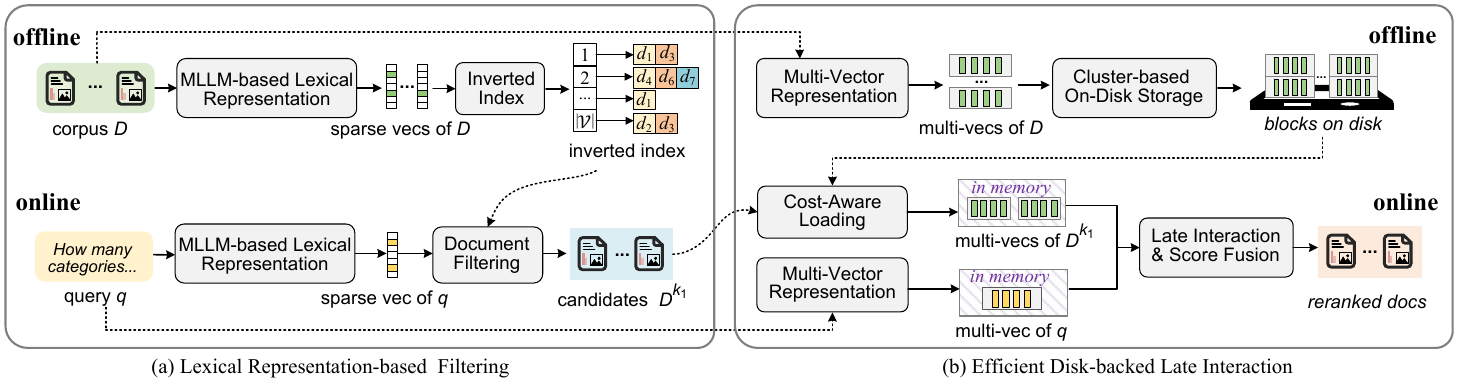}
    \vspace{-6mm}
    \caption{Overview of \textsc{Stellar}. 
   (a) Lexical Representation-based Filtering enables efficient and effective document filtering. (b) Efficient Disk-backed Late Interaction supports low-cost multi-vector similarity computation and sparse-dense score fusion.}
    \label{fig:framework}
    \vspace{-2mm}
\end{figure*}

Figure~\ref{fig:framework} illustrates the overview of \textsc{Stellar}, which comprises two components: lexical representation-based filtering (LRF), and efficient disk-backed late interaction (DLI).

\noindent \textbf{Lexical Representation-based Filtering.}
During the \textit{offline} stage, each document $d$ in the corpus $D$ is encoded into a sparse vector using our proposed MLLM-based sparse lexical representation method. Each sparse vector has a dimensionality equal to the MLLM's vocabulary size $|\mathcal{V}|$, where most dimensions are zero.
These sparse vectors are indexed via an in-memory inverted index that maps each vocabulary dimension (of size $|\mathcal{V}|$) to documents with non-zero values along that dimension, enabling efficient document filtering.
In the \textit{online} phase,the user query $q$ is encoded into a sparse vector. The top-$k_1$ candidate documents $D^{k_1} \subset D$ are filtered by computing sparse similarity scores between the query and document sparse embeddings via the inverted index. 

\noindent \textbf{Efficient Disk-backed Late Interaction.}
During the \textit{offline} stage, each document $d \in D$ is encoded into multiple token embeddings. These token embeddings are stored on disk and organized using the proposed balanced clustering algorithm.
In the \textit{online} phase, we design a cost-aware loading strategy to determine whether to load the entire block or specific token embeddings of the candidate documents $D^{k_1}$ into memory. Meanwhile, the user query $q$ is
encoded into a set of token-level embeddings, which interact with document token embeddings via late interaction to compute multi-vector dense scores. These dense scores are then fused with the first-stage sparse scores to obtain hybrid scores, which are used to rank the candidate documents $D^{k_1}$ as the final results.
\section{Lexical Representation-based Filtering}
\label{sec:lrf}
In this section, we first present the MLLM-based lexical representation method. Then, we describe how to index the sparse vectors and perform document filtering.

\subsection{MLLM-based Lexical Representation} 
\label{subsec:lexical_represent}

We design a siamese dual-encoder built upon an MLLM to encode both documents
and queries into a unified sparse lexical space, which serves as an efficient filtering
mechanism for scalable multimodal document retrieval.

\noindent\textbf{Unified Lexical Projection.} 
We project each document $d \in D$ and text query $q$ into a shared vocabulary space using the MLLM. Specifically, documents and queries are independently processed to extract visual and textual tokens, which are then encoded into last-layer hidden states. Each hidden state is projected into the model's vocabulary space $\mathcal{V}$, yielding a $|\mathcal{V}|$-dimensional lexical vector.
For documents, this projection translates input visual tokens into semantically relevant lexical tokens in $\mathcal{V}$; for text queries, it naturally enables semantic expansion by activating related vocabulary tokens.
To support this unified lexical mapping, we repurpose the MLLM's pre-trained language modeling head $f_{\text{head}}$ as a lexical projection layer. Although originally designed for next-token prediction, $f_{\text{head}}$ learns to map hidden states into a shared vocabulary space during large-scale vision-language pretraining.
We therefore initialize the projection layer with the pre-trained $f_{\text{head}}$ to fully exploit its capability, and further fine-tune it to align with our retrieval objective.

Formally, for each token $\bm{t}_i$ ($i=1, \dots, m$) in a multimodal document $d$, let $\bm{h}_i$ denote its last-layer hidden state. The corresponding lexical vector $\bm{w}_i$ is computed as:
\begin{equation}
\bm{w}_i =  f_{\text{head}}(\bm{h}_i) = [w_{i1}, w_{i2}, \dots, w_{i|\mathcal{V}|}] \in \mathbb{R}^{|\mathcal{V}|}
\end{equation}

\noindent where $w_{ij}$ represents the weight assigned to the $j$-th vocabulary token in representing the token $\bm{t}_i$.
A high value of $w_{ij}$ indicates strong semantic relevance between the token $\bm{t}_i$ and the $j$-th vocabulary token, while a negative value suggests irrelevance and can be set to zero. To implement this, we apply a ReLU function to each vector $\bm{w}_i$, eliminating all negative values.
Finally, max pooling is applied across tokens for each dimension $j$ to extract salient features, yielding the final sparse representation $\textbf{\texttt{enc}}_{\text{spa}}(d) = [z_1, z_2, \dots, z_{|\mathcal{V}|}] \in \mathbb{R}^{|\mathcal{V}|}$, where each $z_j$ is calculated as follows:
\begin{equation}
\label{eq:weight}
     z_j = \max _i \log \left(1+\operatorname{ReLU}\left(w_{i j}\right)\right), \quad i\in\{1, 2, \dots, m\} 
\end{equation}





The sparse lexical representation of a text query $q$, denoted as $\textbf{\texttt{enc}}_{\text{spa}}(q)$, is constructed analogously.

\down\noindent \textbf{MLLM Fine-tuning.} 
High-quality lexical representations are expected to exhibit high similarity between a query and its relevant documents, while maintaining low similarity with irrelevant ones. This enables effective document filtering via similarity search.
To achieve this, we employ contrastive learning~\cite{he2020moco} to fine-tune the MLLM as a sparse encoder $\textbf{\texttt{enc}}_{\text{spa}}$, using in-batch negatives:  
\begin{equation}
\scriptsize
\mathcal{L}_\mathrm{InfoNCE} = -\frac{1}{N} \sum_{i=1}^{N} \log \frac{
    \exp\left(\operatorname{sim}(q_i, d_i^+) / \tau \right)
}{
    \exp\left(\operatorname{sim}(q_i, d_i^+) / \tau \right)
    + \sum\limits_{j\neq i} \exp\left(\operatorname{sim}(q_i, d_j^-) / \tau \right)
}
\end{equation}

\noindent where $\operatorname{sim} (q, d) =  \textbf{\texttt{enc}}_\text{spa}(q) \cdot  \textbf{\texttt{enc}}_\text{spa}(d)$ denotes the similarity between the embeddings of the anchor query $q$ and either the positive (relevant) document $d^+$ or negative (irrelevant) document $d^-$; $N$ denotes the mini-batch size, and $\tau$ is the temperature parameter.

To ensure vector sparsity and improve retrieval efficiency, we incorporate a FLOPs-based regularization loss~\cite{SPLADE} to control the sparsity of the output document representations:
\begin{equation}
\small
\mathcal{L}_{\mathrm{FLOPS}}^d=\sum_{j=1}^{|\mathcal{V}|}\left(\frac{1}{N} \sum_{i=1}^N z_j\left(d_i\right)\right)^2
\end{equation}

\noindent where $z_j\left(d_i\right)$ 
denotes the $j$-th component of the sparse representation for document $d_i$, which 
is computed as shown in Equation (\ref{eq:weight}). The  regularization loss for query representation is denoted as $\mathcal{L}_{\mathrm{FLOPS}}^q$ with a similar definition to $\mathcal{L}_{\mathrm{FLOPS}}^d$. The overall loss is as follows:
\begin{equation}
\small
\mathcal{L}=\mathcal{L}_{\text {InfoNCE}}+\lambda_1 \mathcal{L}^d_{\mathrm{FLOPS}} +
\lambda_2 \mathcal{L}^q_{\mathrm{FLOPS}}
\end{equation}

\noindent where $\lambda_1$ and $\lambda_2$ control the strength of the FLOPs regularization for the document and query vectors, respectively. During fine-tuning, LoRA~\cite{LoRA} is applied to all linear layers in the Transformer blocks and the language modeling head  $f_\text{head}$.

\subsection{Inverted Index and Document Filtering}
\noindent\textbf{Offline Indexing.} 
We use the fine-tuned MLLM as the sparse encoder to generate lexical representations for all the documents $d \in D$. Then, we construct an inverted index over sparse dimensions. Specifically, for each document $d_i \in D$, we record a posting entry $(i, z_j(d_i))$ for each dimension $j \in \mathcal{I}_{d_i}$, where $\mathcal{I}_{d_i}$ denotes the set of non-zero dimensions in its sparse representation, $i$ is the document \textsf{ID}, and $z_j(d_i)$ is the corresponding sparse weight. Posting entries are then grouped by dimension to form the inverted index:
$$
\textsf{Index}[j] = \{(i, z_j(d_i)) \mid z_j(d_i) \neq 0\}, j\in\{1,2,\dots,|\mathcal{V}|\}
$$

\noindent\textbf{Online Filtering.}
At query time, we compute the sparse query representation $\textbf{\texttt{enc}}_{\text{spa}}(q)$ and identify its non-zero dimensions $\mathcal{I}_q$. 
Instead of scanning all documents, we traverse only the posting lists corresponding to dimensions in $\mathcal{I}_q$. 
For each $j \in \mathcal{I}_q$, we access $\textsf{Index}[j]$ and accumulate partial scores for the associated documents.
This process effectively computes the sparse relevance score defined as:
\begin{equation}
\operatorname{score}_\text{spa}(q, d)
= \sum_{j \in \mathcal{I}_q \cap \mathcal{I}_d} z_j(q) \cdot z_j(d),
\end{equation}
where $\mathcal{I}_d$ denotes the set of non-zero dimensions of document $d$.

Based on the sparse scores, we filter the top-$k_1$ ($k_1 \ll |D|$) candidate documents $D^{k_1} \subset D$ and discard the remaining documents from subsequent late-interaction computations.

\section{Disk-backed Late Interaction}
\label{sec:dli}
In this section, we first review token-level multi-vector representation learning, then introduce the cluster-aware on-disk layout design, followed by the cost-aware data loading strategy and score fusion mechanism.

\subsection{Multi-Vector Representation}
To obtain token-level embeddings, we reuse the MLLM backbone in Section~\ref{subsec:lexical_represent} and fine-tune it as a dense multi-vector encoder. Given a multimodal document or a query, we extract the MLLM's last-layer hidden state $\bm{h}_i $ of each visual or text token, and project it into a lower-dimensional embedding space via a linear transformation:
$\bm{e}_i = \mathbf{W}\bm{h}_i + \mathbf{b}$,
where $\mathbf{W}$ and $\mathbf{b}$ denote the projection matrix and bias vector, respectively.
A document and a query are represented as sets of token embeddings, denoted by $\textbf{\texttt{enc}}_{\text{mul}}(d) = \{\bm{e}^d_1,  \dots, \bm{e}^d_m\}$ and
$\textbf{\texttt{enc}}_{\text{mul}}(q) = \{\bm{e}^q_1, \dots, \bm{e}^q_n\}$,
where $m$ and $n$ denote the numbers of visual tokens in document $d$ and text tokens in query $q$, respectively.
We adopt a pairwise contrastive loss~\cite{colpali2025} to fine-tune the MLLM as an effective multi-vector dense encoder:
\begin{equation}
\mathcal{L}_{\text{pw}}=\frac{1}{N} \sum_{i=1}^{N} \log \left(1+\exp \left(\operatorname{s}_i^{-}-\operatorname{s}_i^{+}\right)\right)
\end{equation}

\noindent where $N$ is the number of training examples in each mini-batch, $\operatorname{s}_i^{+} = \operatorname{score}_{\text{mul}}(q_i, d_i^{+})$,
and $\operatorname{s}_i^{-} = \max_{j \neq i} \operatorname{score}_{\text{mul}}(q_i, d_j^{-})$,
both computed using Equation (\ref{eq:late_interaction}). 
We adopt LoRA for fine-tuning.


\subsection{Cluster-based On-Disk Storage} 
We store all learned token-level embeddings of the document corpus on disk rather than in memory, which incurs disk-to-memory loading overhead during online retrieval due to random I/O~\cite{vitter1999io}. At query time, the first-stage (LRF) dynamically filters a set of candidate documents \(D^{k_1}\). If the token embeddings of these filtered retrieved candidate documents are stored contiguously on disk, the cost of random I/O can be substantially reduced. Although the candidate set varies across queries, the following observation enables such an on-disk organization in practice.
Since the first-stage filtering is based on learned lexical representations, the retrieved candidate documents \(D^{k_1}\) tend to exhibit similar sparse embeddings. This property motivates clustering documents by their lexical representations and storing the token embeddings of documents within each cluster contiguously as a \emph{block}. However, naive clustering often produces highly imbalanced block sizes, leading to unpredictable I/O behavior and inefficient data loading (see Section~\ref{subsec:rq4}). To address this, we  design a balanced document clustering algorithm to organize token embeddings on disk.

\begin{algorithm}[!tb]
\DontPrintSemicolon
\small
    \LinesNumbered
    \caption{Balanced Clustering Algorithm (\textsf{BCA})}
    \label{alg:bca}
    
    \KwIn{Sparse vectors $\bm{E}_D$, expected size $s_\text{exp}$, minimum size $s_\text{min}$}
    \KwOut{Balanced clusters $\mathcal{C}_\text{res}$}

    $K \leftarrow \lceil |D|/s_\text{exp} \rceil, \:\:  \mathcal{C}_\text{tmp} \leftarrow \emptyset$\\
    $\mathcal{C}_\text{init} \leftarrow k\text{-means}(\bm{E}_D, K)$\\
    
    \For{each cluster $c \in \mathcal{C}_\text{init}$}{
        \If{$|c| > s_\text{exp}$}{
            $\mathcal{C}_\text{tmp} \leftarrow \mathcal{C}_\text{tmp} \cup \textsf{RecursiveSplit}(c, s_\text{exp})$
        }
        \Else{
            $\mathcal{C}_\text{tmp} \leftarrow \mathcal{C}_\text{tmp} \cup \{c\}$
        }
    }

    $\mathcal{C}_{\text{small}} \leftarrow \{c \in \mathcal{C}_{\text{tmp}} \mid |c| < s_{\text{min}}\}$\\
    $\mathcal{C}_{\text{surv}} \leftarrow \mathcal{C}_{\text{tmp}} \setminus \mathcal{C}_{\text{small}}$\\
    
    \For{each document $d \in \bigcup_{c \in \mathcal{C}_{\text{small}}} c$}{
        $c_{\text{target}} \leftarrow \operatorname{argmax}_{c \in \mathcal{C}_{\text{surv}}} \text{sim}(d, \text{centroid}(c))$\\
        Move $d$ to $c_{\text{target}}$
    }
    \Return $\mathcal{C}_{\text{res}} \leftarrow \mathcal{C}_{\text{surv}}$

    \SetKwProg{Fn}{Procedure}{:}{}
    \Fn{\textnormal{\textsf{RecursiveSplit}($c$, $s_\text{exp}$)}}{
        $n_k \leftarrow \lceil |c| / s_\text{exp} \rceil, \:\:
        \mathcal{C}_{\text{out}} \leftarrow \emptyset$\;
        $\mathcal{C}_{\text{sub}} \leftarrow k\text{-means}(c, n_k)$\;
        \For{each $s \in \mathcal{C}_{\text{sub}}$}{
            \lIf{$|s| > s_\text{exp}$}{$\mathcal{C}_{\text{out}} \leftarrow \mathcal{C}_{\text{out}} \cup \textsf{RecursiveSplit}(s, s_\text{exp})$}
            \lElse{$\mathcal{C}_{\text{out}} \leftarrow \mathcal{C}_{\text{out}} \cup \{s\}$}
        }
        \KwRet $\mathcal{C}_{\text{out}}$
    }
\end{algorithm}

\noindent \textbf{Balanced Clustering Algorithm ($\textsf{BCA}$).}
The algorithm proceeds in three stages, with pseudocode shown in Algorithm~\ref{alg:bca}.

\noindent{{\textit{\underline{Initial Partitioning.}}}} \textsf{BCA} first sets the initial number of clusters to $K=\lceil |D|/s_{\text{exp}} \rceil$ and performs a preliminary $k$-means partitioning (lines 1–2), where $s_{\text{exp}}$ denotes the target cluster capacity. 

\noindent{{\textit{\underline{Recursive Splitting.}}}}
Then, \textsf{BCA} identifies clusters whose sizes exceed the capacity $s_{\text{exp}}$ and triggers a recursive splitting strategy (lines 3--5). Within the \textsf{RecursiveSplit} procedure (lines 14--20), the local cluster number $n_k$ is dynamically recomputed based on the current cluster size. This hierarchical refinement continues until all sub-clusters are size-compliant, ensuring that no single disk access loads an oversized block that would stall I/O.

\noindent{{\textit{\underline{Locality-Aware Reassignment.}}}}
To avoid overly small clusters, we define a minimum cluster size $s_{\text{min}}$. Clusters with fewer than $s_{\text{min}}$ documents are classified as $\mathcal{C}_{\text{small}}$ and dissolved (lines 8–9). Their constituent documents are then reassigned to the clusters in the surviving set $\mathcal{C}_{\text{surv}}$ whose centroids exhibit the highest similarity to the document embeddings (lines 10--12). This reassignment ensures that each final cluster in $\mathcal{C}_{\text{res}}$ maintains sufficient document density, maximizing the effective utilization of each disk access and reducing overall random I/O overhead.
After balanced clustering, the token-level embeddings of documents belonging to the same cluster are stored contiguously on disk, forming a storage block $b$.



\noindent \textbf{Two-level Index.}
To efficiently manage and load on-disk token embeddings, we design two lightweight in-memory indexes.

\noindent{{\textit{\underline{Document-level index}}:}}
$\textsf{ID}_d \mapsto (\textsf{ID}_b, \texttt{count}, \texttt{offset},   \texttt{len})$, where $\textsf{ID}_d$ is a globally unique document identifier; $\textsf{ID}_b$ denotes the block containing the document's token embeddings; $\texttt{count}$ specifies the number of token vectors for document $d$; $\texttt{offset}$ indicates the starting byte offset of these embeddings within the block; and $\texttt{len}$ denotes their total byte length.

\noindent{\textit{\underline{Block-level index}}:}
$\textsf{ID}_b \mapsto (\texttt{doc\_list}, \texttt{total})$, where $\textsf{ID}_b$ denotes the block identifier, \texttt{doc\_list} is the list of document IDs contained in the block, and  \texttt{total}  indicates the number of vectors in the block.





\subsection{Cost-Aware Loading}
\label{subsec:loading}
During the online stage, we aim to fetch the token embeddings of candidate documents $D^{k_1} =\{d_1,  \dots, d_{k_1}\}$ from disk.
We first identify the set of \emph{hit blocks} $\mathcal{B}^*=\{b_i^*\}$, each containing embeddings of at least one candidate document. For each hit block $b_i^*$, we consider two data loading modes:

\noindent{{\textit{\underline{(i) Full Block Loading (\textsf{FBL})}}}}: 
This mode leverages high sequential I/O throughput by loading the entire block into memory in a single continuous read. Subsequently, the block-level index is used to locate and retain only the vector slices corresponding to the candidate documents. 

\noindent{{\textit{\underline{ (ii) Specific Vector Loading  (\textsf{SVL})}}}}: 
This mode eliminates the overhead of irrelevant data loading by utilizing the document-level index to directly locate and load embeddings of candidate documents within a hit block. 
\textsf{FBL} benefits from high sequential I/O throughput but incurs extra overhead from loading embeddings of non-candidate documents, making it inefficient when candidates occupy only a small fraction of a block.
\textsf{SVL} is efficient when the hit block contains only a small number of candidate documents, but its efficiency is constrained by the low throughput of non-contiguous random disk accesses.
The key challenge is thus to determine the optimal loading model for each hit block during online stage.
 
We design a cost-aware loading strategy that selects between \textsf{FBL} and \textsf{SVL} based on an estimated I/O cost. The estimation is guided by a simple yet effective cost model, whose key parameters include the vector dimension $V_{\text{dim}}$, the byte size per float $B_{\text{float}}$, and the disk's sequential and random read rates, denoted by $R_{\text{seq}}$ and $R_{\text{rand}}$, respectively.
To measure $R_{\text{seq}}$, we sequentially read a temporary binary file (e.g., 1 GB) and compute the throughput as the total bytes read divided by the elapsed time. To estimate $R_{\text{rand}}$, we perform a large number of small random reads (e.g., 10,000 reads of 100 KB) from the same file and compute the average throughput.
For each hit block $b_i^* \in \mathcal{B}^*$, we obtain the total number of vectors $N_{i,\text{total}}$ in the block and the number of required vectors $N_{i,\text{req}}$ for the candidate documents using the block-level index. We then estimate the loading time for \textsf{FBL} and \textsf{SVL} using the following cost model, and choose the strategy with the lower estimated cost.

\begin{equation}
\label{eq:T_block}
   T_\textsf{FBL} = \frac{N_{i,\text{total}} \times V_\text{dim} \times B_\text{float}}{R_\text{seq}} 
\end{equation}
\begin{equation}
\label{eq:T_vec}
    T_\textsf{SVL} = \frac{N_{i,\text{req}}  \times V_\text{dim} \times B_\text{float}}{R_\text{rand}}  
\end{equation}


\noindent \textbf{Discussion.} 
Although \textsf{FBL} incurs higher per-iteration memory usage in Algorithm~\ref{alg:bca}, it does not affect \textsc{Stellar}'s peak memory footprint.
Specifically, our objective is to load token embeddings for $k_1$ documents. 
Under \textsf{FBL}, approximately $s_\text{exp}$ document's embeddings are loaded per block, where $s_\text{exp}$ denotes the expected cluster size defined in Algorithm~\ref{alg:bca}.
Crucially, $s_\text{exp}$ is configured to be much smaller than $k_1$; otherwise, loading just a single block would exceed the total number of required documents, resulting in inefficient memory usage. Therefore, the peak memory usage is dominated by the late interaction stage rather than the data loading stage. 
The primary difference in memory usage between \textsf{FBL} and \textsf{SVL} arises from index storage:
\textsf{FBL} maintains both block-level and document-level indexes in memory, while \textsf{SVL} only requires the document-level index.
However, these indexes are lightweight and introduce negligible memory overhead (see Section~\ref{subsec:rq4}).
Consequently, memory usage is excluded from our cost model.

\subsection{Late Interaction and Score Fusion}
After disk-to-memory data loading, we obtain all the token embeddings of candidate documents $D^{k_1}$. The user query is also encoded into multiple token embeddings, followed by late interaction with each document $d_i \in D^{k_1}$ using Equation (\ref{eq:late_interaction}) to obtain the multi-vector dense scores $\operatorname{score}_\text{mul}(q, d_i)$.

Instead of relying solely on dense scores to rank candidate documents, we fuse them with the sparse score $\operatorname{score}_\text{spa}(q, d_i)$ obtained from the first-stage filtering, as the two scores provide complementary retrieval signals~\cite{Pneuma}. 
The fused score is defined below:
\begin{equation}
\small
\label{eq:fusion}
    \operatorname{score}_\text{fus}(q, d_i) = \alpha \cdot  {Z}(\operatorname{score}_\text{spa}(q, d_i)) +  {Z}(\operatorname{score}_\text{mul}(q, d_i))  
\end{equation}

\noindent where $\alpha$ is a coefficient between 0 and 1, and the function $ {Z}(\cdot)$ denotes the $z-$normalization, i.e., $ {Z}(x) = \frac{x-\mu}{\delta}$ where $\mu$ is the mean value and $\delta$ is the standard deviation.
We use the fused $\operatorname{score}_\text{fus}$ to rank candidate documents in $D^{k_1}$ to obtain the final results.


\section{Experiments}
\label{sec:exp}
In this section, we evaluate \textsc{Stellar} on four existing benchmark datasets and a newly proposed large-scale dataset LargeDoc.
The code and datasets are publicly available at
\url{https://github.com/ZJU-DAILY/Stellar}.

\begin{table}[t]
\small
\centering
\caption{Statistics of the datasets used in experiments.}
\vspace{-0.1in}
\renewcommand{\arraystretch}{1}
\resizebox{1\linewidth}{!}{
\begin{tabular}{cccccc} 
\toprule
\textbf{Scale} & \textbf{Dataset} & \textbf{Document Type} & \textbf{Train} & \multicolumn{2}{c}{\textbf{Evaluation}} \\ 
\cmidrule(lr){5-6}
& & & \textbf{\# $(q, d)$} & \textbf{\# $q$} & \textbf{\# $d$} \\ 
\midrule
\multirow{2}{*}{Small} 
 & DocVQA & Industrial Docs & 10,624 & 591 & 741 \\
 & InfoVQA & Infographics & 17,664 & 718 & 459 \\
\midrule
\multirow{2}{*}{Medium} 
 & ArXivQA & ArXiv Figures & 25,856 & 816 & 8,066 \\
 & PlotQA & Scientific Plots & 56,192 & 863 & 9,593 \\
\midrule
Large 
 & LargeDoc & Mixed & --- & 94 & 400,000 \\
\bottomrule
\end{tabular}}
\label{tab:statistic}
\vspace{-5mm}
\end{table}

\subsection{Experimental Setup}
\noindent{\textbf{Datasets.}}
We adopt four existing benchmark datasets and propose a large-scale dataset. The datasets with statistics are summarized in Table~\ref{tab:statistic}.

\noindent \underline{\textit{Existing benchmark datasets.}}
We adopt four widely-used benchmark datasets in document retrieval and open-domain QA tasks~\cite{VisRAG2025}:
(1) \textbf{DocVQA}~\cite{MPDocVQA2023}, consisting of scanned industrial documents with dense textual layouts, complex forms, and tabular structures;  (2) \textbf{InfoVQA}~\cite{infographicvqa22}, composed of visually-rich infographics with complex spatial layouts;  (3) \textbf{ArXivQA}~\cite{arxivqa2024}, 
containing scientific figures extracted from academic papers;
and (4) \textbf{PlotQA}~\cite{PlotQA20}, comprising scientific plots with structured axes and markers.

\noindent \underline{\textit{Large-scale dataset.}}
To the best of our knowledge, there is no publicly available large-scale dataset for multimodal document retrieval.
To fill this gap, we construct \textbf{LargeDoc}, a large-scale benchmark consisting of 400,000 multimodal documents and 94 test queries, each paired with a ground-truth document. LargeDoc is constructed based on Docmatix~\cite{docmatix}, which contains 2.4 million PDFs spanning diverse types such as papers, forms, charts, and slides, and 9.5 million questions paired with related documents.
The construction process comprises three steps.
(1) Query selection. Since many Docmatix questions are designed for closed-domain QA rather than retrieval (e.g., ``What is this page about?''), we manually curate 94 high-quality query-document pairs suitable for retrieval evaluation.
(2) Corpus scaling. To emulate a realistic large-scale retrieval setting, we randomly sample over 400,000 documents from Docmatix and merge the 94 ground-truth documents into this pool.
(3) False-negative cleaning. Large-scale retrieval benchmarks often suffer from false negatives --- unlabeled documents that also satisfy a query.
To mitigate this issue, we apply state-of-the-art text-based and vision-based retrievers (gte-Qwen2, DSE, and ColPali) to retrieve the top-100 candidates for each query, and take the union of all retrieved candidates.
These candidates are then verified by GPT-4o, followed by manual validation from three post-graduate students.
Any document confirmed by at least two annotators as a valid alternative to the ground-truth document is removed from the pool.
Finally, we remove excess irrelevant documents to limit the corpus size to 400,000 documents.

\vspace{1mm}
\noindent \textbf{Baselines.} 
We compare \textsc{Stellar} with three categories of multimodal document retrieval methods. 
For fairness, all methods perform \textit{exact similarity search} except for \textsf{QColPali}.

\noindent \underline{\textit{Text-centric methods.}} These methods extract text from document pages using PP-OCR~\cite{paddleocr} and perform text retrieval using text retrieval methods.

\begin{itemize}[leftmargin=*]
\item \textsf{{BM25}}~\cite{bm2509}: A classic sparse retrieval method that evaluates document relevance by measuring term frequency and statistical significance within the collection.

\item  \textsf{{BGE-Large}}~\cite{bgelarge2023}: A powerful dense encoder that maps textual content into a high-dimensional vector space to facilitate semantic similarity matching. 

\item \textsf{{gte-Qwen2}}~\cite{nvembed2025}: An LLM-based embedding model released by Alibaba, built upon the same \textsf{Qwen2-1.5B} backbone as the MLLM \textsf{Qwen2-VL-2B}.
\end{itemize}

\noindent \underline{\textit{Vision-centric single-vector methods:}}
These methods treat each multimodal document page as an image, and encode each query and document page into a single vector via MLLMs.

\begin{itemize}[leftmargin=*]
\item \textsf{{DSE}}~\cite{DSE24}: A method appends a special token to the input sequence of MLLMs and uses the final hidden state of this token as dense representations.

\item \textsf{{VisRet}}~\cite{VisRAG2025}: A method uses position-weighted mean pooling over the final hidden states of MLLMs to get dense representations

\item \textsf{{Bi-Qwen}}: A method performs mean pooling over the final hidden states of MLLMs to obtain dense representations.
\end{itemize}
 
\noindent \underline{\textit{Vision-centric multi-vector methods:}}
Methods in this category encode each document page and query into multiple token
embeddings, enabling fine-grained late interaction.

\begin{itemize}[leftmargin=*]
\item \textsf{{ColPali}}\cite{colpali2025}: The state-of-the-art multi-vector multimodal document retrieval method which performs exact late interaction of query-document token embeddings~\cite{ColBERT2020}.

\item \textsf{{QColPali}}: An approximate variant of \textsf{ColPali} that incorporates the \textsc{PLAID}~\cite{PLAID} method. Each token embedding is represented by a centroid and a quantized residual vector. Following \textsc{PLAID}, \textsf{QColPali} first filters $k_1$ candidate documents using centroid embeddings and then reconstructs the quantized token embeddings for late interaction.
\end{itemize}

\begin{table*}[t]
\small
\centering
\vspace{-2mm}
\caption{Retrieval performance across six datasets, evaluated with Recall@10 (R@1), Recall@10 (R@10), and MRR@10 (M@10).}
\vspace{-0.1in}
\label{tab:hyper_retrieval}
\renewcommand{\arraystretch}{1.2}
\setlength{\tabcolsep}{1.2mm}{

\begin{tabular}{l|ccc|ccc|ccc|ccc|ccc}
\toprule
\multirow{2}{*}{\textbf{Methods}}
 
& \multicolumn{3}{c|}{\textbf{DocVQA}}
& \multicolumn{3}{c|}{\textbf{InfoVQA}}
& \multicolumn{3}{c|}{\textbf{ArXivQA}}
& \multicolumn{3}{c|}{\textbf{PlotQA}} 
& \multicolumn{3}{c}{\textbf{LargeDoc}} \\
& {R@1} & {R@10} & {M@10}
& {R@1} & {R@10} & {M@10}
& {R@1} & {R@10} & {M@10}
& {R@1} & {R@10} & {M@10}
& {R@1} & {R@10} & {M@10} \\
\hline
\textsf{BM25}
& 60.74 & 86.80 & 75.27
& 45.41  & 82.59 & 66.94
& 30.15 & 54.29 & 43.65
& 30.93 & 76.01 & 57.28
& 57.45 & 70.21 & 62.77
\\
 
\textsf{BGE-Large}
& 45.57 & 68.19 & 50.76
& 58.39 & 88.16 & 72.38
& 28.23 & 48.65 & 39.29
& 28.46 & 73.12 & 51.33 
& 60.64 & 73.40 & 65.21 
\\
\textsf{gte-Qwen2}
& 59.89 & 83.25 & 68.11
& 64.76 & 92.83 & 78.69
& 32.84 & 58.14 & 43.12
& 31.05 & 78.66 & 53.69 
& 62.77 & 75.53 & 69.05
\\
\hline
\textsf{DSE }
& 63.96 & 90.19 & 72.79
& 76.46 & 96.38 & 84.02
& 67.28 & 84.44 & 72.71
& 31.87 & 73.46 & 44.00 
& 71.28 & 80.85 & 73.74 
\\
\textsf{VisRet }
& 61.25 & 87.64 & 70.79
& 75.77 & 96.10 & 83.08
& 64.83 & 82.72 & 70.52
& 30.48 & 72.89 & 42.22 
& 68.09 & 77.66 & 71.28
\\
\textsf{Bi-Qwen}
& 61.59 & 88.49 & 71.05
& 75.77 & 96.10 & 83.65
& 64.09 & 82.84 & 70.16
& 26.88 & 69.29 & 39.14 
& 67.02 & 78.72 & 70.14 
\\
\hline


\textsf{ColPali}  
&\underline{82.74} & \underline{97.27} & \underline{88.26}
&\underline{86.49} & \underline{98.89} & \underline{91.61}
&\underline{75.74} & \underline{88.48} & \underline{80.02}
&\underline{56.55} & \textbf{87.02} & \textbf{66.27} 
&\underline{74.47} & \underline{86.17} & \underline{77.30} 
\\
\textsf{QColPali}

&80.03 & 96.62 & 86.53
&83.71 & 98.33 & 89.71
&73.65 & 87.01 & 78.14
&55.50 & 87.02 & 65.88 
&72.34 & 85.11 & 75.89 
\\

\textsc{Stellar} (\textbf{ours})
&\textbf{84.43} & \textbf{97.29} & \textbf{89.05}
&\textbf{86.91} & \textbf{99.16} & \textbf{92.10}
&\textbf{75.75} & \textbf{88.97} & \textbf{80.24}
&\textbf{56.90} & \underline{86.10} & \underline{66.01} 
&\textbf{75.53} & \textbf{87.23} & \textbf{79.47} 
\\


\bottomrule
\end{tabular}}
\vspace{-2mm}
\end{table*}

\begin{table*}[t]
\centering
\caption{Peak memory usage (MB) and query latency (ms) of the vision-centric multi-vector methods.}
\small
\vspace{-0.1in}
\label{tab:mem_time_extended}
\renewcommand{\arraystretch}{1}
\setlength{\tabcolsep}{7pt}
\begin{tabular}{l|cc|cc|cc|cc|cc}
\toprule
\multirow{3}{*}{\textbf{Methods}} 
  & \multicolumn{4}{c|}{ Small Scale } 
  & \multicolumn{4}{c|}{ Medium Scale } 
  & \multicolumn{2}{c}{ Large Scale } \\
\cmidrule(lr){2-11}
 
  & \multicolumn{2}{c|}{DocVQA} 
 
  & \multicolumn{2}{c|}{InfoVQA} 
  & \multicolumn{2}{c|}{ArXivQA} 
  & \multicolumn{2}{c|}{PlotQA} 
  & \multicolumn{2}{c}{LargeDoc} \\
  & memory & latency
  & memory & latency
  & memory & latency
  & memory & latency
  & memory & latency \\
\midrule
\textsf{ColPali} 
   & 315.90 & 73.44 &  183.76 & 59.80 
  & 3,301.28 & 649.35 & 3,765.12 & 518.49 & 147,355 & 42,115 \\

\textsf{QColPali}
   & 40.00 & \textbf{45.81} &  \textbf{25.60} & \textbf{42.39} 
  & 398.40 & 261.59 & 480.00 & 302.43 & 13,699 & 613  \\
  
\textsc{Stellar} (\textbf{ours}) 
 & \textbf{32.51} & 50.16 &  38.66 & 44.73 
  & \textbf{49.56} & \textbf{42.35} & \textbf{45.27} & \textbf{43.02} & \textbf{988} & \textbf{110}   \\
\bottomrule
\end{tabular}
\vspace{-2mm}
\end{table*}

\noindent\textbf{Evaluation metrics.}
To evaluate the effectiveness, we follow the previous study~\cite{VisRAG2025} to adopt Recall-at-k (R$@k$) aggregated over the test queries, and Mean Reciprocal Rank-at-k (M$@k$).
We also report peak memory usage and query latency to evaluate the scalability of the proposed method.

\vspace{1mm}
\noindent\textbf{Implementation Details.}
All vision-centric methods use \textsf{Qwen2-VL-2B}~\cite{qwen2vl2024} as the backbone, which consists of a Vision Transformer and the \textsf{Qwen2-1.5B} language model. To ensure a fair evaluation, the backbone model is trained for a single epoch on the same mixed training set, which combines all benchmarks' training splits and another corpus provided by the previous study~\cite{VisRAG2025}. 
We adopt LoRA~\cite{LoRA} for fine-tuning, with rank $r=32$. The temperature $\tau$ is set to 1. Following insights from the previous study~\cite{SPLADE}, we set the FLOPs regularization strengths to 9e-5 and 6e-5 for documents and queries, respectively. 
We filter $k_1 = 100$ candidate documents during the first stage.
In our balanced clustering algorithm, we empirically set the expected cluster size $s_\text{exp}$ to 50 and the minimum cluster size $s_\text{min}$ to 3. 
All experiments were conducted on a server equipped with an Intel(R) Xeon(R) Silver 4316 CPU (2.30GHz), 256GB RAM.
The online retrieval process is performed solely on the CPU.

\subsection{Overall Performance }

\noindent \textbf{Effectiveness.} Table~\ref{tab:hyper_retrieval} reports the R@1, R@10, and M@10 of \textsc{Stellar} and other baselines. Results in \textbf{bold} are the best; \underline{underlined} are the second. We have several key observations:
\begin{itemize}
\item Vision-centric methods consistently outperform text-centric methods on all benchmarks, which is consistent with the observations from previous studies~\cite {VisRAG2025, colpali2025}.
\item Wthin the vision-centric category, single-vector methods underperform multi-vector methods by an average of 15 points in R@1, underscoring the limitation of representing an entire document with a single global vector.
\item Compared with existing multi-vector methods, \textsc{Stellar} achieves the best R@1 and consistently high R@10 and M@10 across all datasets. 
Its accuracy is comparable to, and even outperforms \textsf{ColPali}, which performs inefficient late interaction between query and all documents.
Its superior accuracy arises from (1) lexical representation-based filtering (see Section~\ref{subsec:rq3}), and (2) the score fusion mechanism that integrates both sparse and dense signals to rank the candidate documents (see Section~\ref{subsec:rq4}).
In contrast, \textsf{QColPali} reduces memory via quantization, resulting in inevitable effectiveness degradation.
\end{itemize}

\down \noindent \textbf{Memory and Latency.}
Table~\ref{tab:mem_time_extended} reports the memory usage and query latency of vision-centric \textit{multi-vector} methods across datasets of different scales. 
We observe that on small-scale datasets, i.e., DocVQA and InfoVQA, \textsf{QColPali} achieves memory and latency reductions comparable to \textsc{Stellar}. However, as the corpus size increases, the limitations of the baselines become apparent.  On medium-scale datasets ArXivQA and PlotQA, \textsf{ColPali} and \textsf{QColPali} require 74.5$\times$ and 9.3$\times$ the memory of \textsc{Stellar}, respectively. 
On LargeDoc, \textsc{Stellar} requires only \textbf{0.96 GB} of memory, achieving a \textbf{150$\times$} reduction compared to \textsf{ColPali} and a \textbf{14$\times$} reduction relative to \textsf{QColPali}.

\begin{figure}[t]
    \centering
    \includegraphics[width=0.98\linewidth]{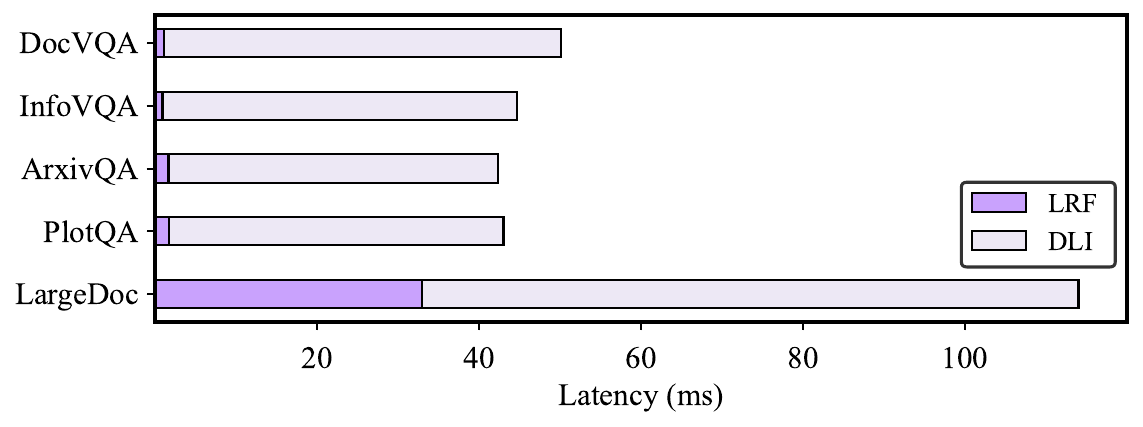}
    \vspace{-3mm}
    {\footnotesize
        \caption{Runtime of lexical representation-based filtering (LRF) and disk-backed late interaction (DLI).}
        \label{fig:latency_stacked}
    }
\end{figure}

\begin{figure}[t]
    \centering
    \includegraphics[width=0.98\linewidth]{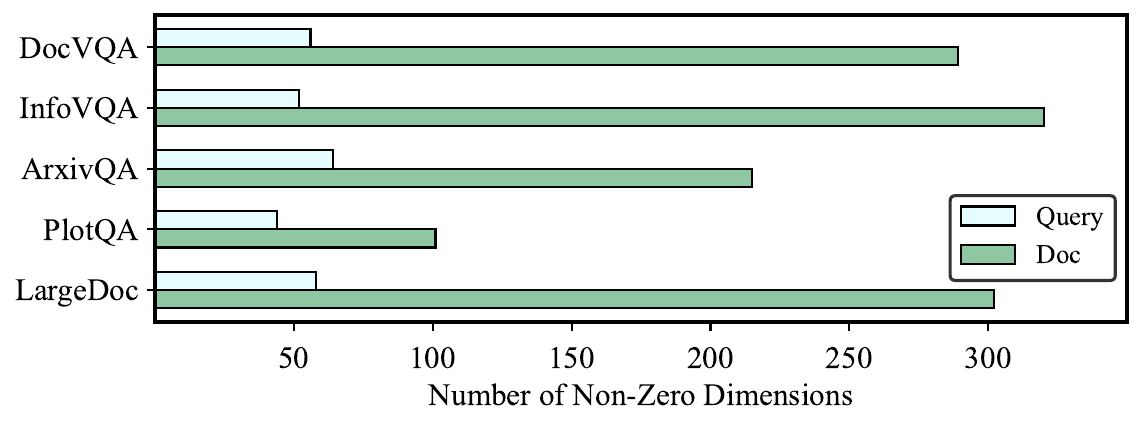}
   
    \caption{Average number of non-zero dimensions of our sparse representations (vocabulary dimension \(\lvert \mathcal{V}\rvert = 151{,}936\)).}
    \label{fig:activation_dimensions}
    \vspace{-4mm}
\end{figure}

We conduct an in-depth analysis of the runtime of the two phases in \textsc{Stellar}, as illustrated in 
Figure~\ref{fig:latency_stacked}. The first observation is that the LRF stage exhibits extremely low latency, attributed to the high sparsity of the learned representations. Specifically, our learned sparse representation reduces the original 151,936-dimensional vocabulary-based vectors to sparse vectors with only around 200 active (non-zero) dimensions on average, achieving a sparsity rate of approximately 99.87\%, as shown in 
Figure~\ref{fig:activation_dimensions}. 
This high sparsity not only minimizes memory usage but also enables the use of inverted index structures for fast large-scale candidate filtering.
The second observation is the DLI stage's latency depends on both the dataset and its scale. Generally speaking, larger datasets require processing more hit blocks and thus incur higher latency. Note that DocVQA contains high-resolution document images, leading to more visual tokens and consequently higher late-interaction latency.

\subsection{Scalability Study } 
\label{subsec:scale}
\begin{figure}[t] 
    \centering
        \includegraphics[width=0.6\linewidth]{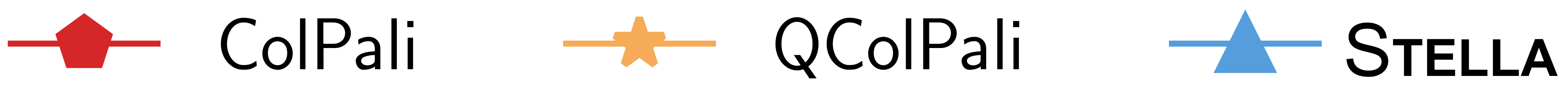}\\
    \begin{minipage}[t]{0.45\linewidth}
        \centering
        \includegraphics[width=\linewidth]{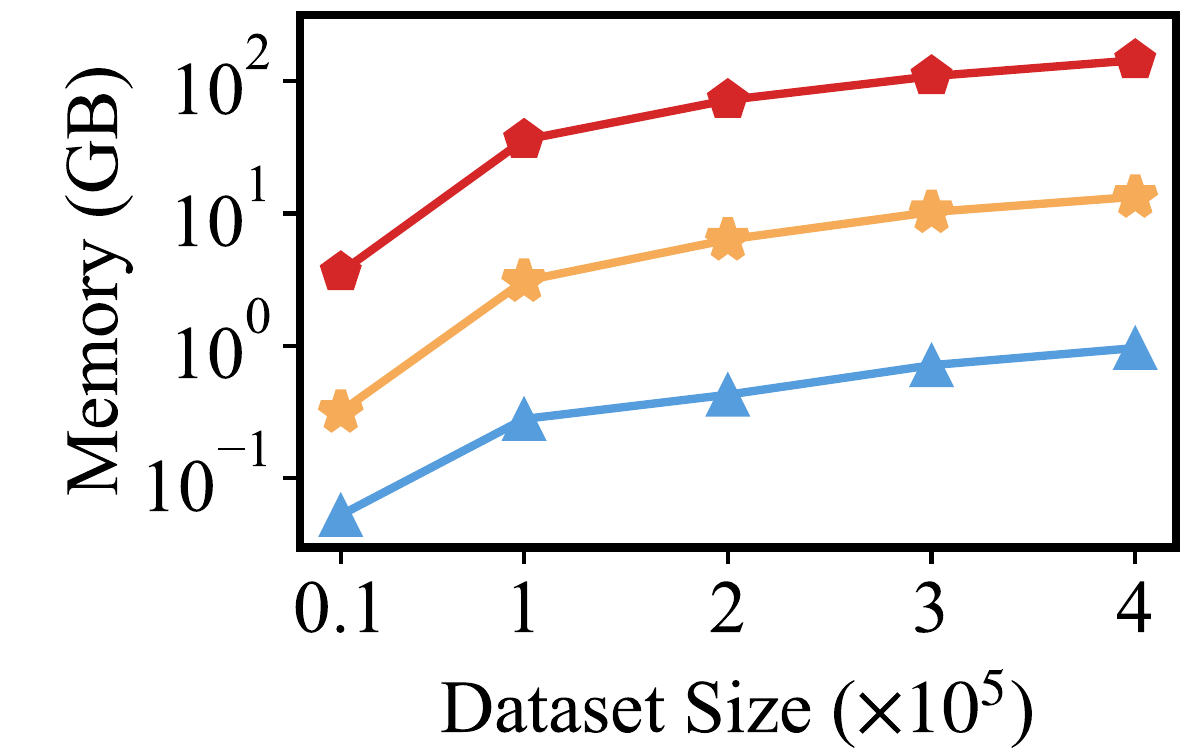}
        \vspace{1mm}
        {\footnotesize (a) Memory w.r.t different scales}
    \end{minipage}%
    \hspace{2mm}
    \begin{minipage}[t]{0.45\linewidth}
        \centering
        \includegraphics[width=\linewidth]{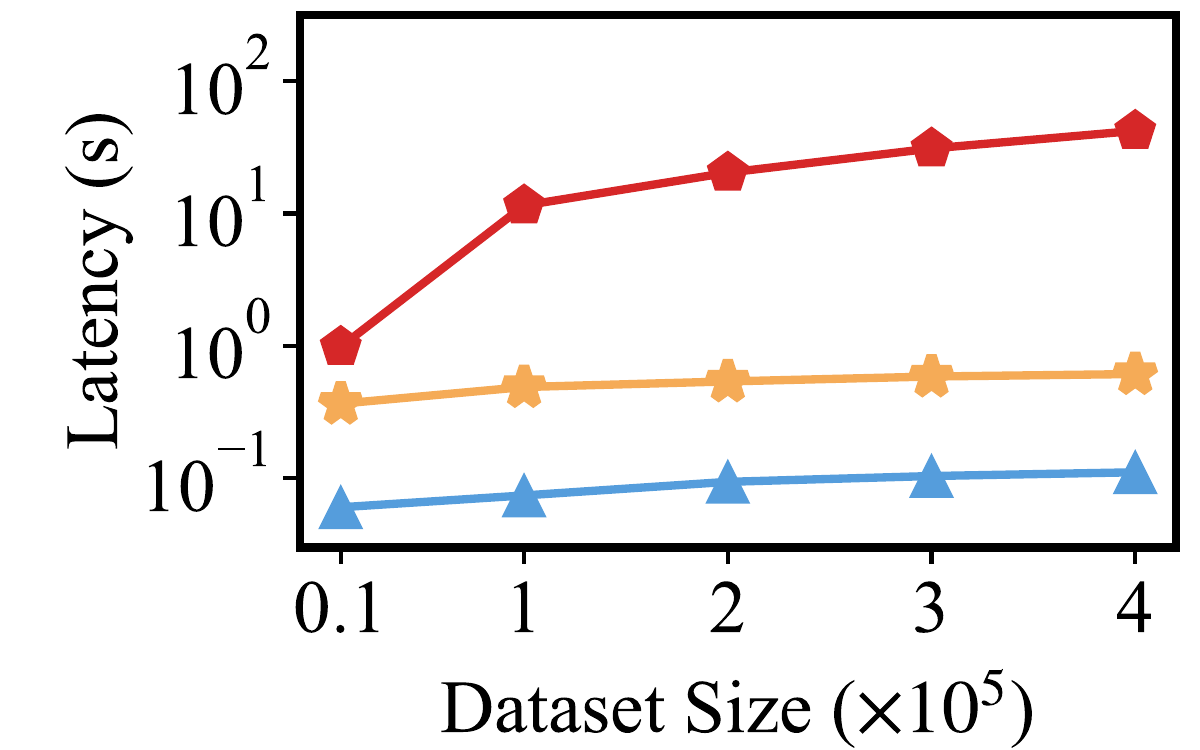}
        \vspace{1mm}
        {\footnotesize (b) Latency w.r.t different scales}
    \end{minipage}
 
    \caption{Scalability study with different dataset size.}
    \label{fig:scalability}
    \vspace{-2mm}
\end{figure}

We measure the memory usage and query latency of the vision-centric multi-vector methods on LargeDoc at varying scales.
For memory usage, as shown in Figure~\ref{fig:scalability}(a), 
\textsf{ColPali} exhibits the largest memory footprint. As the corpus scale increases, its memory consumption grows most rapidly, resulting in more than 150 GB when reaching 400K documents. 
Although \textsf{QColPali} compresses high-dimensional token embeddings, its memory usage still grows rapidly with corpus size. 
Moreover, it results in memory overflow on our 256 GB RAM machine when the dataset exceeds 400K documents, primarily due to the costly compression stage.   
In contrast, \textsc{Stellar} maintains a low memory footprint, reducing memory usage by \textbf{1-2} orders of magnitude compared with \textsf{ColPali} and \textsf{QColPali}. Moreover, 
its memory consumption grows more than \textbf{10$\times$} more slowly with increasing corpus size than \textsf{QColPali}, highlighting its superior scalability.
For query latency, 
as shown in Figure~\ref{fig:scalability}(b), \textsf{ColPali} is 2 orders of magnitude slower than the other two methods due to its exhaustive late interaction across all documents. In contrast, both \textsf{QColPali} and \textsc{Stellar} limit late interaction to the top-$k_1$ candidates. However, \textsc{Stellar} achieves higher efficiency due to the efficient LRF stage.

\begin{figure} 
    \centering
  
        \includegraphics[width=0.6\linewidth]{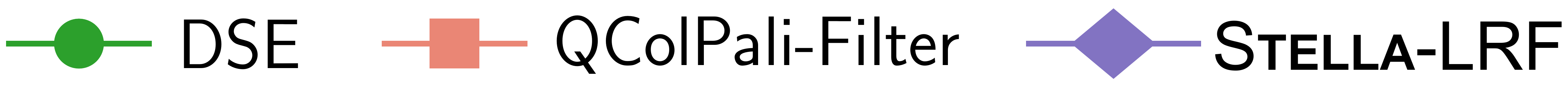}
    \\
    
    \begin{minipage}[t]{0.45\linewidth}
        \centering
        \includegraphics[width=\linewidth]{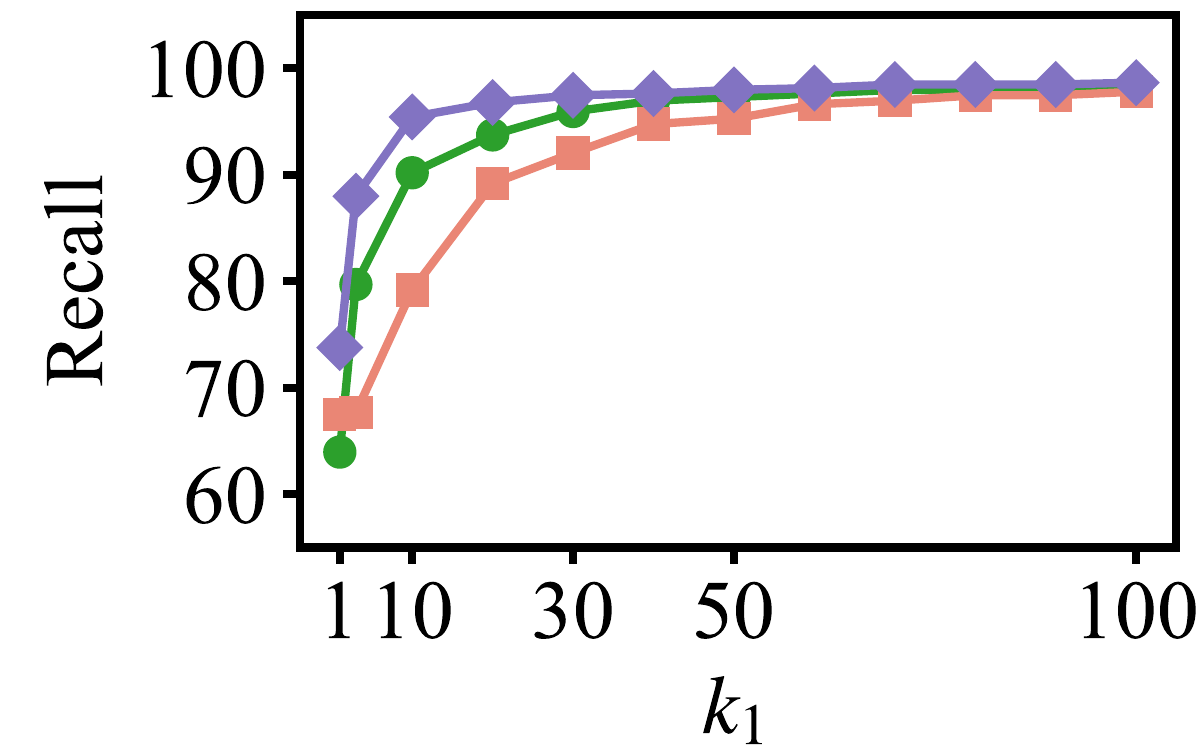}
        \vspace{1mm}
        {\footnotesize (a) Recall@$k_1$ on DocVQA}
    \end{minipage}%
    \hspace{2mm}
    \begin{minipage}[t]{0.45\linewidth}
        \centering
        \includegraphics[width=\linewidth]{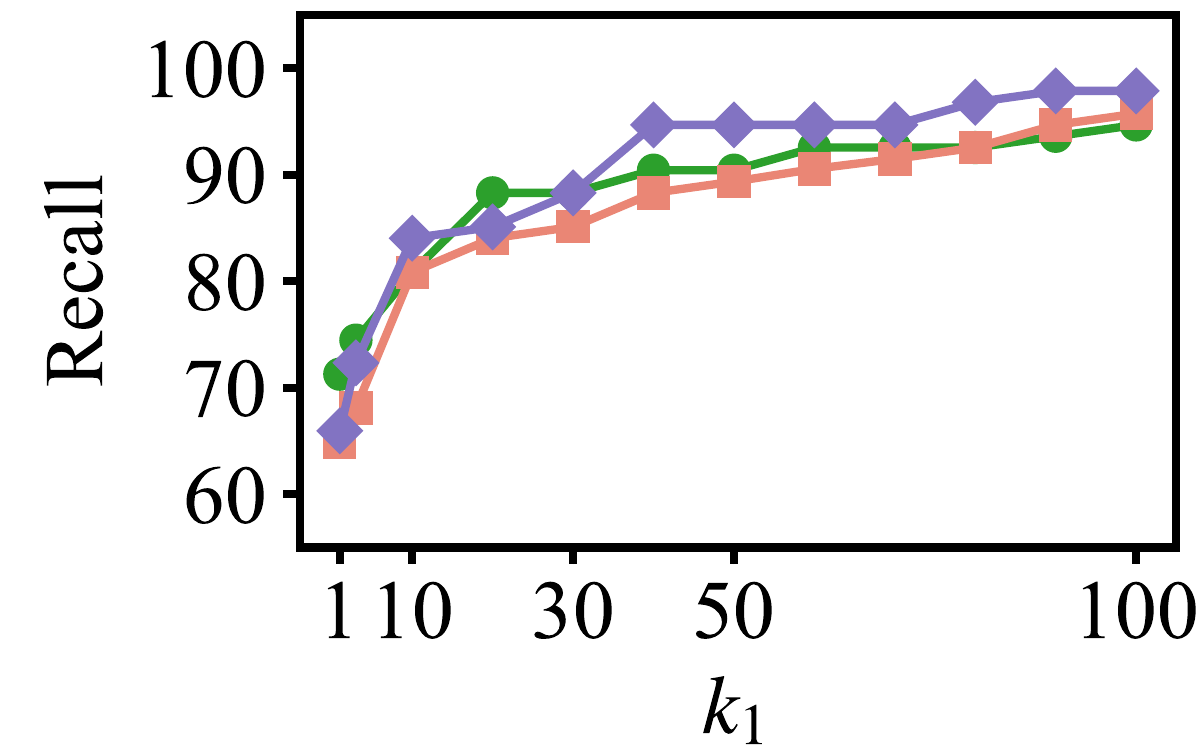}
        \vspace{1mm}
        {\footnotesize(b) Recall@$k_1$ on LargeDoc}
    \end{minipage}
    
        \includegraphics[width=0.6\linewidth]{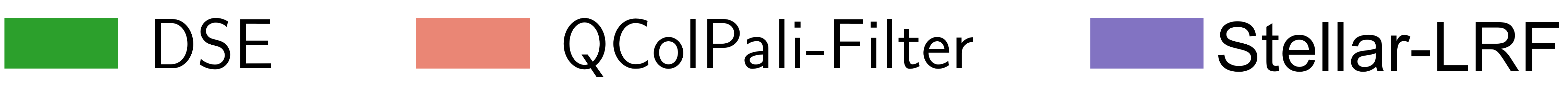}\\
 
    \begin{minipage}[t]{0.45\linewidth}
        \centering
        \includegraphics[width=\linewidth]{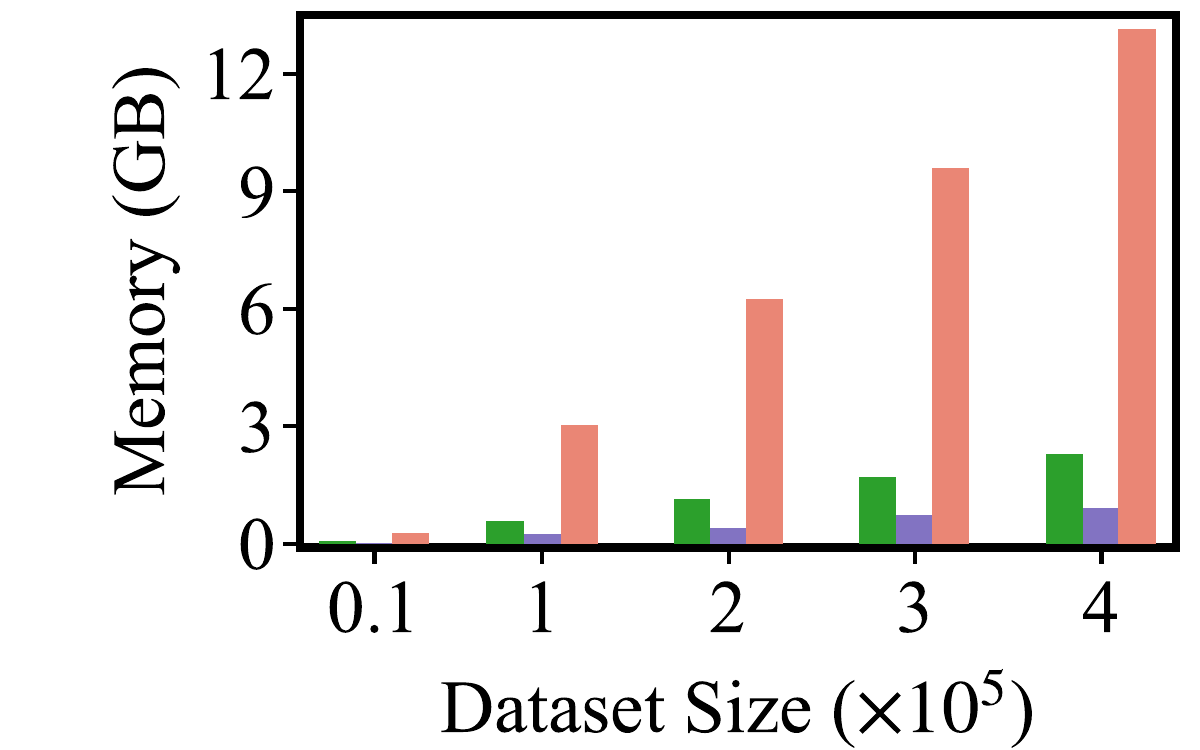}
        {\footnotesize (c) Memory usage on LargeDoc}
    \end{minipage}
    \hspace{2mm}
    \begin{minipage}[t]{0.45\linewidth}
        \centering
        \includegraphics[width=\linewidth]{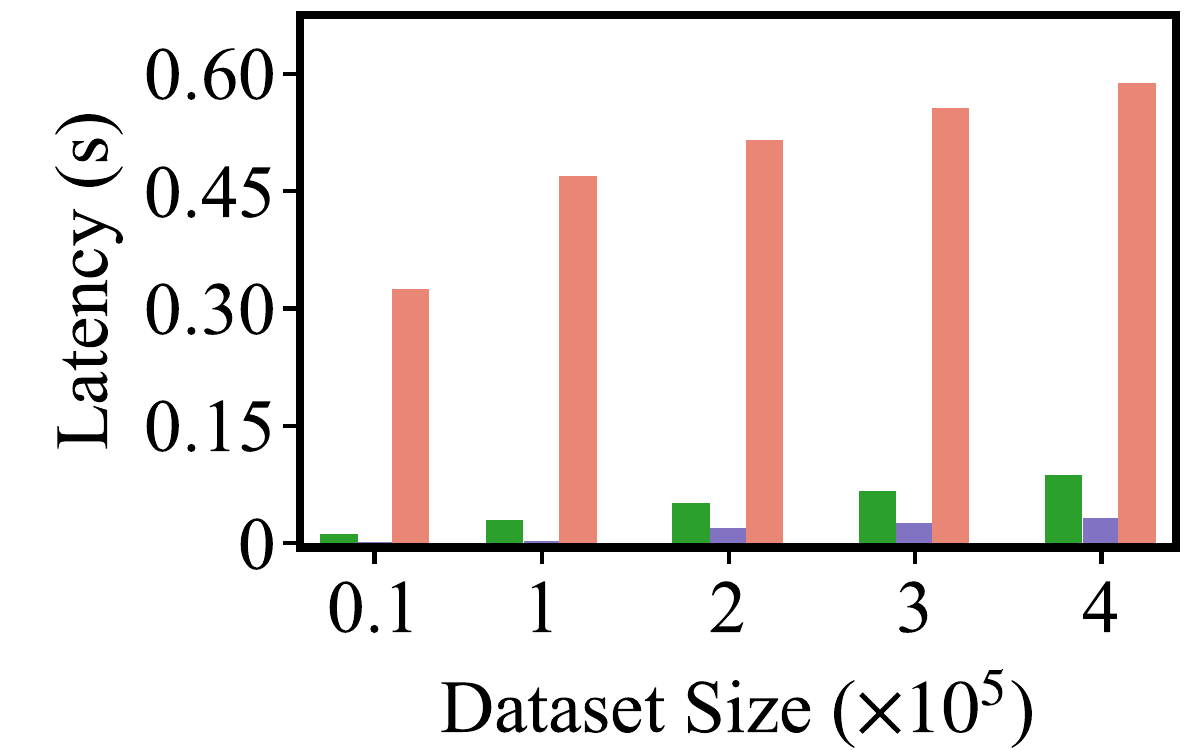}
        {\footnotesize (d) Runtime on LargeDoc}
    \end{minipage}
 
    \caption{Performance of different filtering methods.} 
    \label{fig:LRF}
    \vspace{-2mm}
\end{figure}

\subsection{Effectiveness of LRF }
\label{subsec:rq3}

We compare our lexical representation-based filtering (\textsc{Stellar}-LRF) with two alternative filtering methods: (i) \textsf{DSE}, the strongest dense single-vector retriever (see Table~\ref{tab:hyper_retrieval}), which serves as a dense first-stage filter; and (ii) \textsf{QColPali-Filter}, which filters candidates using centroids obtained by clustering token-level embeddings.

We show Recall@$k_1$ across different $k_1$ on a small-scale dataset and a large-scale dataset in Figure~\ref{fig:LRF}(a)-(b); other datasets exhibit similar trends. 
Notably, our LRF consistently achieves higher recall than the two dense-vector-based filtering methods, demonstrating the effectiveness of lexical representations in capturing lexical cues often overlooked by dense embeddings.
When $k_1$ reaches 100, all three methods attain stable and high recall. However, LRF is more efficient in both time and memory.
As shown in Figure~\ref{fig:LRF}(c)-(d), on the LargeDoc, LRF exhibits lower memory usage and runtime.
It is worth noting that on the 400K-scale, using dense \textsf{DSE} for first-stage filtering requires  2.3 GB of memory---more than twice the end-to-end memory consumption of \textsc{Stellar} (0.96 GB; see Table~\ref{tab:mem_time_extended}). This indicates that \textit{dense filtering can become a major memory bottleneck}, underscoring the necessity of our sparse lexical filtering.

\begin{figure}  
    \centering
    \begin{minipage}[t]{0.45\linewidth}
        \centering
        \includegraphics[width=\linewidth]{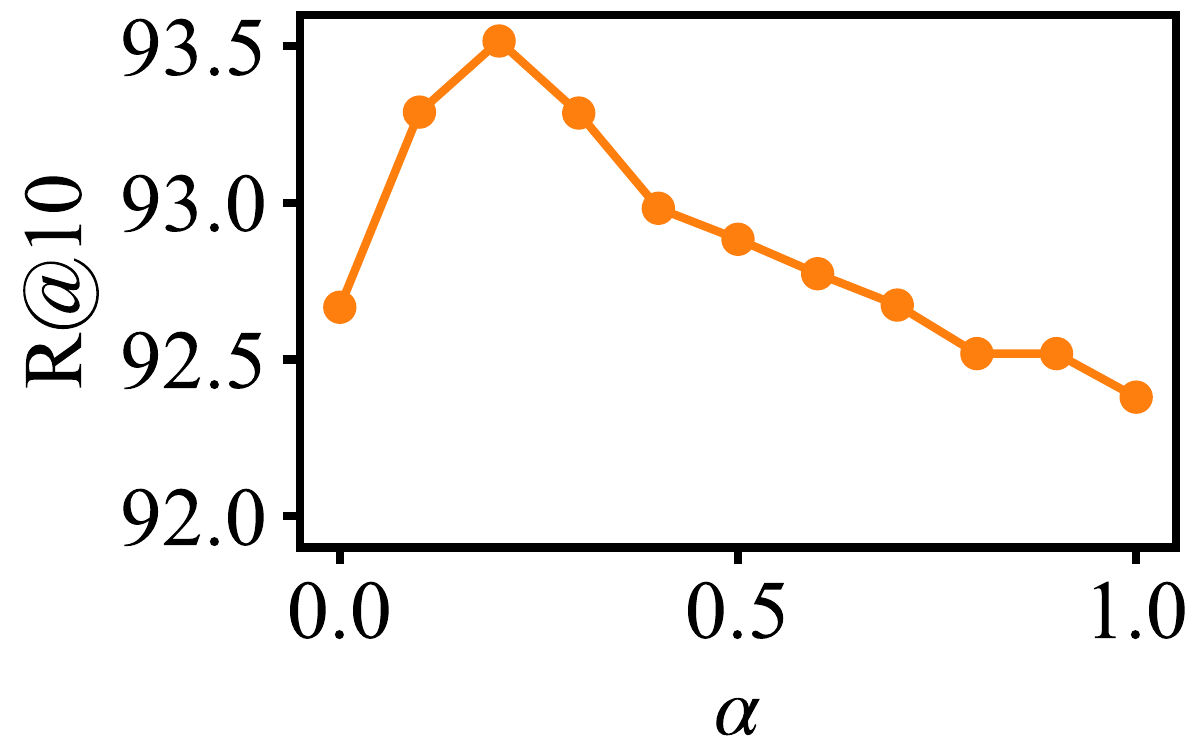}
        \vspace{1mm}
        {\footnotesize (a) Effect of $\alpha$ on average R@10}
    \end{minipage}%
    \hspace{2mm}
    \begin{minipage}[t]{0.45\linewidth}
        \centering
        \includegraphics[width=\linewidth]{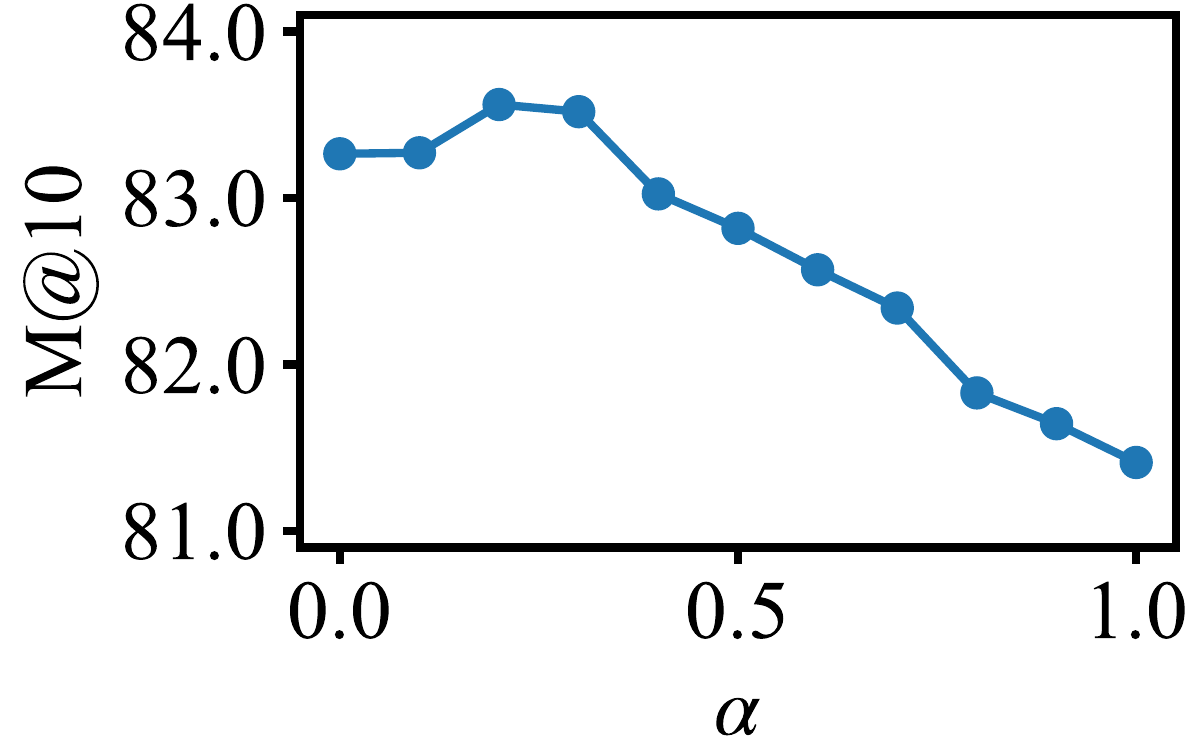}
        \vspace{1mm}
        {\footnotesize (b) Effect of $\alpha$ on average M@10}
    \end{minipage}
    \caption{Sensitivity analysis of $\alpha$ in score fusion.}
     \vspace{-3mm}
    \label{fig: sensitivity}
\end{figure}

\begin{table*}[t] 
\centering
\small
\caption{Memory usage (MB) and latency (ms) with different document clustering and data loading strategies.}
\small
\vspace{-0.1in}
\label{tab:loading_ablation}
\renewcommand{\arraystretch}{1.1}
 \setlength{\tabcolsep}{7.5pt}
\begin{tabular}{l|cc|cc|cc|cc|cc}
\toprule
\multirow{2}{*}{\textbf{Methods}} 
  & \multicolumn{2}{c|}{DocVQA} 
  & \multicolumn{2}{c|}{InfoVQA} 
  & \multicolumn{2}{c|}{ArXivQA} 
  & \multicolumn{2}{c|}{PlotQA}
  & \multicolumn{2}{c}{LargeDoc} \\
  & memory & latency
  & memory & latency
  & memory & latency
  & memory & latency
  & memory & latency \\
\midrule
\textsc{Stellar}  
 
& 32.51 & \textbf{50.16} 
 
& 38.66 & \textbf{44.73} 
& 49.56 & \textbf{42.35} 
& 45.27 & \textbf{43.02}
& \textbf{988} & \textbf{110} \\ 
\hline
\textsf{NCA}
 
& 33.11 & 60.33 
 
& 38.63 & 51.06 
& 50.21 & 55.81  
& 45.81 & 57.45
& 1014  &  133 \\

\textsf{FBL} 
 
& 34.60 & 178.11 
 
& 39.31 & 189.22 
& 52.51 & 158.65  
& 45.89 & 176.59
& 1002 & 351 \\ 

\textsf{SVL} 
 
& \textbf{31.02} & 63.20 
 
& \textbf{38.35} & 60.68 
& \textbf{48.63} & 64.47 
& \textbf{44.53} & 59.47
& 989 & 142 \\
\bottomrule
\end{tabular} 
\vspace{-1mm}
\end{table*}

\begin{table*}  
\centering
\caption{Performance Comparison of  \textsc{Stellar}  and \textsc{Stellar} w/o Score Fusion (SF).}
\vspace{-0.1in}
\small
\label{tab:ablation_acc}
\renewcommand{\arraystretch}{1.2}
\setlength{\tabcolsep}{2pt}{
\begin{tabular}{l|ccc|ccc|ccc|ccc|ccc}
\specialrule{.08em}{.06em}{.06em}
\multirow{2}{*}{\textbf{Methods}}
 
& \multicolumn{3}{c|}{DocVQA}
 
& \multicolumn{3}{c|}{InfoVQA}
& \multicolumn{3}{c|}{ArXivQA}
& \multicolumn{3}{c|}{PlotQA} 
& \multicolumn{3}{c}{LargeDoc}
\\
 
& {R@1} & {R@10} & {M@10}
& {R@1} & {R@10} & {M@10}
& {R@1} & {R@10} & {M@10}
& {R@1} & {R@10} & {M@10}
& {R@1} & {R@10} & {M@10}
\\
\specialrule{.05em}{.06em}{.06em}

\textsc{Stellar} 
 
& \textbf{84.43} & \textbf{97.29} & \textbf{89.05}
& \textbf{86.91} & \textbf{99.16} & \textbf{92.10}
& \textbf{75.75} & \textbf{88.97} & \textbf{80.24}
& \textbf{56.90} & 86.10 & \textbf{66.01}
& \textbf{75.53} & 87.23 & \textbf{79.47}
\\

\textsc{Stellar} w/o SF
& 82.91 & 96.95 & 88.25
& 86.49 & 98.89 & 91.63
& 75.49 & 88.73 & 79.86
& 56.20 & \textbf{86.44} & 65.80
& 70.21 & \textbf{88.30} & 75.69 
\\

\specialrule{.08em}{.06em}{.06em}
\end{tabular} 
}
\vspace{-2mm}
\end{table*}

\subsection{Impact of Strategies in DLI}
 \label{subsec:rq4}
We conduct an ablation study to validate the necessity of our designs in disk-backed late interaction (DLI), focusing on: (1) the effect of balanced clustering for on-disk layout optimization on memory and latency, (2) the effect of the selective loading strategy, and (3) the effectiveness of score fusion.

\down\noindent \textbf{Impact of Clustering Strategy.}
We replace our balanced document clustering algorithm with a naive clustering algorithm (\textsf{NCA}), which performs clustering without adjusting cluster sizes. As shown in Table~\ref{tab:loading_ablation}, \textsf{NCA} results in consistently higher latency than \textsc{Stellar} across all datasets, highlighting the importance of well-balanced clusters for
efficient disk-to-memory data loading.

\down\noindent \textbf{Impact of Loading Strategy.}
We report the memory usage and query latency by replacing our cost-aware loading strategy with (1) full block loading (\textsf{FBL}), and (2) specific vector loading (\textsf{SVL}).
Table~\ref{tab:loading_ablation} reports the results. The first observation is that the memory costs are similar across loading strategies, consistent with our discussion in Section~\ref{subsec:loading}.
The second observation is that both \textsf{FBL} and \textsf{SVL} incur higher latency than our method, highlighting the effectiveness of our cost model in dynamically selecting the optimal strategy.

\down\noindent \textbf{Impact of Score Fusion.}
We remove the score fusion (\textsf{SF}) component and rank documents solely based on the dense late-interaction scores following first-stage filtering, as shown in Table~\ref{tab:ablation_acc}. The results demonstrate the complementary nature of sparse and dense signals, highlighting the importance of score fusion in enhancing accuracy.
To balance sparse and dense scores, we introduce a weighting parameter $\alpha$. As shown in  Figure~\ref{fig: sensitivity}, sensitivity analysis shows that accuracy first improves with increasing $\alpha$, peaks at 0.3, then declines, offering useful guidance for tuning $\alpha$ in practice.
\section{Conclusion}
\label{sec:conclusion}
In this work, we present \textsc{Stellar},
a scalable framework for multimodal document retrieval. 
 Instead of compressing the token embeddings of large document collections
to fit in memory, \textsc{Stellar} stores them on disk. We introduce
a lexical representation method that encodes both multimodal documents and text queries into a unified lexical space,
enabling effective and efficient candidate filtering. To mitigate
disk I/O overhead, we propose a balanced clustering–based
storage optimization method, along with a cost-aware online loading strategy for efficient retrieval. Comprehensive experiments demonstrate that \textsc{Stellar} substantially reduces
memory usage and latency while maintaining high retrieval
effectiveness. In the future, we plan to further accelerate disk-to-memory loading through hardware-aware prefetching or lightweight embedding caching, and explore multi-document retrieval in more complex scenarios.



 
%

\bibliographystyle{IEEEtran}
\bibliography{IEEEabrv,sample-base}

\vfill

\end{document}